\documentclass[traditabstract,a4]{aa}  

\usepackage{epsfig}
\usepackage{graphicx}
\usepackage{epstopdf}
\usepackage{color}  
\usepackage{array}   
\usepackage{units}  
\usepackage[breaklinks, colorlinks, citecolor=blue]{hyperref}
\usepackage{natbib}  
\usepackage{multirow}   
\usepackage{amsmath,amssymb}  
\usepackage[english]{babel}
\usepackage[latin1]{inputenc}
\usepackage[T1]{fontenc}
\usepackage{longtable,lscape}
\usepackage{footnote}

\usepackage{txfonts}
\usepackage[toc,page]{appendix} 
\begin{document}

%%%%%%%%%%%%%

%%%%%%%%%%%%%%%
\title{Modeling the cross power spectrum of Sunyaev-Zel'dovich and X-ray surveys}
\author{G.Hurier}
   \author{G. Hurier\inst{1}
   \and N. Aghanim\inst{1}
   \and M. Douspis\inst{1}
          }

\institute{Institut d'Astrophysique Spatiale, CNRS (UMR8617) and Universit\'{e} Paris-Sud 11, B\^{a}timent 121, 91405 Orsay, France \\
\email{ghurier@ias.u-psud.fr} 
}

   \date{Received /Accepted}
 
   \abstract{Thermal Sunyaev-Zel'dovich (tSZ) effect and X-ray emission from galaxy clusters have been extensively used to constrain cosmological parameters. These constraints are highly sensitive to the relations between cluster masses and observables (tSZ and X-ray fluxes).
 The cross-correlation of tSZ and X-ray data is thus a powerful tool, in addition of tSZ and X-ray based analysis, to test our modeling of both tSZ and X-ray emission from galaxy clusters.
      We chose to explore this cross correlation as both emissions trace the hot gas in galaxy clusters and thus constitute one the easiest correlation that can be studied.\\
   We present a complete modeling of the cross correlation between tSZ effect and X-ray emission from galaxy clusters, and focuses on the dependencies with clusters scaling laws and cosmological parameters.\\
   We show that the present knowledge of cosmological parameters and scaling laws parameters leads to an uncertainties of 47\% on the overall normalization of the tSZ-X cross correlation power spectrum.\\
We present the expected signal-to-noise ratio for the tSZ-X cross-correlation angular power spectrum considering the sensitivity of actual tSZ and X-ray surveys from {\it Planck}-like data and ROSAT. 
We demonstrate that this signal-to-noise can reach 31.5 in realistic situation, leading to a constraint on the amplitude of tSZ-X cross correlation up to 3.2\%, fifteen times better than actual modeling limitations. 
Consequently, used in addition to other probes of cosmological parameters and scaling relations, we show that the tSZ-X is a powerful probe to constrain scaling relations and cosmological parameters.}


   \maketitle
 
\section{Introduction}

Galaxy clusters are the largest virialized structures, they can be observed through X-ray, via the bremsstrahlung emission produced by the ionized intra-cluster medium \citep[see e.g.][]{boh00,ebe00,ebe01}. This hot intra-cluster medium also produces a distortion of the CMB black-body emission via the thermal Sunyaev-Zel'dovich (tSZ) effect \citep{sun69,sun72}. This effect was observed toward a large number of clusters by Planck \citep{planckESZ,PlanckPSZ}, ACT \citep{mar11} and SPT \citep{rei13}.\\
The number of galaxy clusters is extremely sensitive to cosmological parameters, especially to the normalization of the matter power spectrum, $\sigma_8$, and to the matter density, $\Omega_{\rm m}$.\\
It is thus possible to use galaxy cluster catalogs to constraint cosmological parameters \citep{van10,seh11,PlanckSZC} through a halo mass function formalism.\\

We have now access to a full sky coverage for both X-ray emission with the ROSAT all sky survey (RASS), and tSZ emission with {\it Planck} \citep{planckSZS}.
Consequently, beyond tSZ clusters catalogs analysis, it is possible to perform tSZ angular power spectrum analysis.
This process allows to consider all clusters on the covered sky without any selection function \citep[see e.g.,][]{planckSZS}, contrary to catalog based analysis. This allows to catch the signals from higher redshift and lower mass objects that are not detected individually. Such measurement is limited by the contamination produced by other astrophysical components, mainly the cosmic infra-red background \citep[CIB,][]{pug96,fix98}.\\
It is difficult to perform the same power spectrum analysis with X-ray surveys. The X-ray photons, at low energy (< 0.5 keV), are absorbed by neutral hydrogen of our Galaxy and, at higher energy, the X-ray sky power is dominated by the emission from Active Galactic Nuclei (AGN). Consequently, X-ray surveys are most commonly used to constraint the AGN spatial clustering \citep{kru10,miy11,kru12}.\\

In addition to auto-correlation power spectrum analysis, the cross power spectrum between tSZ effect and X-ray emission can be used. 
This cross-correlation is one of the easiest correlation to study, as both signals are produced by the same hot gas of electrons. 
Using such approach allows to minimize the contamination by other astrophysical components and suppresses the instrumental noise contribution to the power spectrum.\\

tSZ-X cross spectrum is sensitive to both X-ray and tSZ scaling relations \citep[see e.g,][for present constraints on scaling relations]{ben13,planckSL,arn10}. This sensitivity limits the determination of cosmological parameters. However, it offers another possibility to constraint tSZ and X-ray scaling laws.\\ 

The utilization of the tSZ-X correlation has already been discussed in the literature.
 \citet{die03} has attempted to directly compare theoretical prediction with the measured cross power spectrum between WMAP temperature anisotropies maps and ROSAT All Sky Survey (RASS). 
 The limited sensitivity and resolution of the WMAP experiment leads to upper limits on the tSZ-X correlation.\\ 
 More recently, \citep{haj13} performed the measurement of the cross correlation between the tSZ sky and an X-ray based catalog of clusters.
 From their analysis they derive $\sigma_8 (\Omega_{\rm m}/0.30)^{0.26} = 0.80 \pm 0.02$.\\

We present in this paper, an up-to-date modeling of the tSZ/X-ray cross-correlation. 
In Sect.~\ref{modelxsz}, we present our modeling of the tSZ-X cross correlation. 
We give a particular attention to the distribution in mass and redshift of the tSZ-X power. In Section~\ref{secres}, we explore the variations of the tSZ-X spectrum with respect to cosmological and scaling laws parameters. We also discuss modeling uncertainties considering our knowledge on cosmological and scaling law parameters and we present the main limitations for the tSZ-X correlation measurement using simulated {\it Planck}-like data.
Finally in Sect.\ref{secpred}, we predict the expected signal-to-noise for the tSZ-X correlation from simulations of {\it Planck}-like tSZ survey and ROSAT All Sky Survey, and we present  the associated constraints on cosmological and scaling law parameters.\\

Throughout the paper, we used the {\it Planck}-CMB best fitting cosmology \citep{planckPAR} as our fiducial cosmological model, unless otherwise specified. Thus, we consider $H_0 = 67.1 \pm 1.4$ km/s/Mpc, $\sigma_8 = 0.834 \pm 0.027$ and $\Omega_{\rm m} = 0.317 \pm 0.020$.

\section{Modeling tSZ-Xray cross correlation}
\label{modelxsz}

\begin{figure*}[!th]
\begin{center}
\includegraphics[scale=0.4]{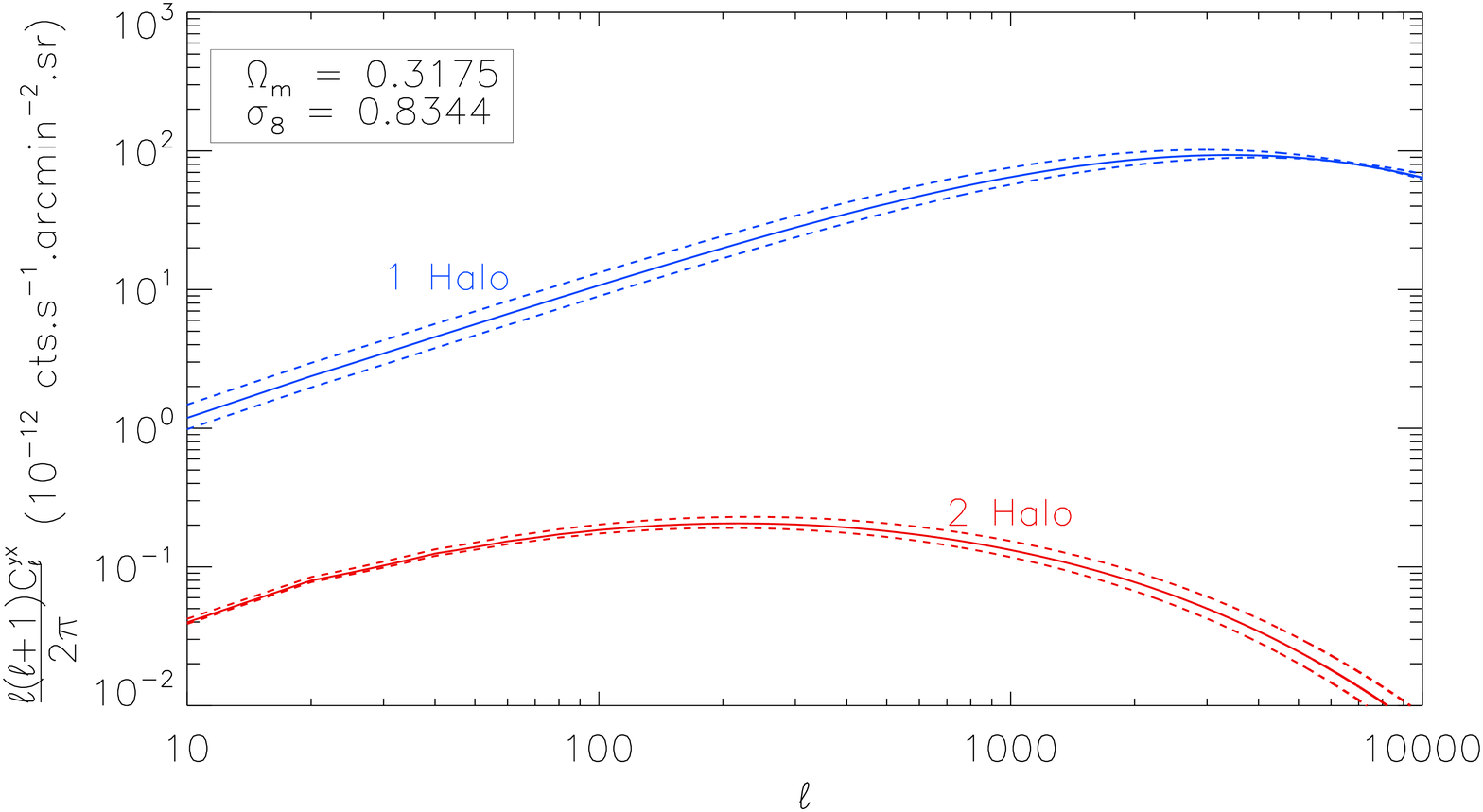}
\caption{tSZ-Xray cross correlation for the 1-halo term in blue and for the 2-halo term in red, considering the ROSAT hard band from 0.5 to 2.0 keV. The dashed lines represent the 1 $\sigma$ level considering variations of scaling relations power laws. This cross correlation spectrum have been predicted using our fiducial cosmological model.}
\label{crossxsz}
\end{center}
\end{figure*}
\subsection{The tSZ effect from galaxy clusters}

The tSZ effect consists of a small spectral distortion of the CMB black-body \citep{sun69,sun72}, its intensity is related to the integral of the pressure across the line-of-sight via the Compton parameter. This parameter in a given direction of the sky reads 
\begin{equation}
y = \int \frac{k_{\rm B} \sigma_{\rm T}}{m_{\rm e} c^2} n_{\rm e} T_{\rm e} {\rm d}l,
\end{equation} 
where ${\rm d}l$ is the distance along the line-of-sight, $n_{\rm e}$ and $T_{\rm e}$ are the electron number density and the temperature, respectively.\\
In units of CMB temperature, the contribution of the tSZ effect to the sub-millimeter sky intensity for a given observation frequency $\nu$ is given by $\frac{\Delta T_{\rm CMB}}{T_{\rm CMB}} = g(\nu) y$.
Neglecting relativistic corrections we have $g(\nu) = \left[ x {\rm coth} \left(\frac{x}{2}\right) - 4\right]$, with $x = h\nu/(k_{\rm B} T_{\rm CMB})$. This function is equal to 0 around $217$~GHz, it is negative at lower frequencies and positive for higher frequencies. Thus the spectral distortion induced by the hot gas of baryons provides a characteristic signal allowing to directly measure the pressure distribution in galaxy clusters. 
In the context of a $\Lambda$CDM cosmology, this spectral distortion is known to be independent of the redshift. This have been tested and validated for a redshift range from 0 to 1 \citep{hur13b}.
This characteristic spectral distortion can be used to separate the tSZ from other emissions of the microwave sky to derive Compton parameter map \citep[see e.g,][]{rem11,hur13a,planckSZS}.\\

\subsection{The X-ray emission from galaxy clusters}
\label{secbrem}

The ionized gas in the intra-cluster medium produces an X-ray emission via Bremsstrahlung. This radiation is proportional to the square of the electronic density. 
The energy spectrum of the X-ray emission from galaxy clusters depends mainly on the temperature, $T_{500}$, of the intra-clusters medium and to a lesser extent on the metallicity, $Z$, of the gas.
To model the metallicity evolution history, we follow \citet{and12} using the relation derived from the analysis of 130 galaxy clusters in a redshift range from 0.1 to 1.3,
\begin{equation}
\overline{Z} = \frac{0.35}{1-{\rm exp}(-11/6)}(1-{\rm exp}(-t(z)/6)),
\end{equation}
with $t(z)$ the age of the universe at a redshift $z$.\\
From an observational point of view, the X-ray spectrum depends on the redshift, $z$.
X-ray photons are also absorbed by the neutral hydrogen in our Galaxy. This absorption is particularly significant for photons with $E < 0.5$~keV. Consequently, The observed count-rate depends on the column density of neutral hydrogen, $n_{\rm H}$, on the line-of-sight.\\
To estimate the X-ray flux from each cluster, we compute an unabsorbed X-ray spectrum, $\phi_{\rm unabs} = {\rm d}n_\gamma/{\rm d}E$, with $n_\gamma$ the number of emitted photons by the cluster inside a radius of $R_{500}$, using a \textsc{MEKAL} model \citep{mew85}. %for a given temperature and a given metal abundance. 
To do this, we use a relation between the physical properties of the cluster, mass and redshift, and the temperature, such relations are presented in Sect.~\ref{secscal}. 
Then, we computed the absorbed spectrum,$\phi_{\rm abs}$, as
\begin{equation}
\phi_{\rm abs}(E) = \phi_{\rm unabs}(E)\, {\rm exp}\left[-n_{\rm H} \sigma(E)\right],
\end{equation}
with $\sigma(E)$ the photoelectric cross-section. The unabsorbed luminosity, $L_{500}$, in a given energy bin $[E_{\rm min},E_{\rm max}]$ of a cluster reads
\begin{equation}
L_{500} = \int_{E_{\rm min}}^{E_{\rm max}} {\rm d}E\, E\, \phi_{\rm abs}(E)
\end{equation} 
Finally, the expected number count in a given energy bin $[E'_{\rm min},E'_{\rm max}]$ is computed as 
\begin{equation}
S_{500} = \frac{a}{4 \pi \chi^2(z)}\int_{E'_{\rm min}}^{E'_{\rm max}} {\rm d}E' \int_{0}^{\infty} {\rm d}E\, {\cal M}(E',E) \,A(E)\, \phi_{\rm abs}(E),
\end{equation}
where $\chi(z)$ is the comoving angular distance at redshift $z$, $a$ the scale factor, $A(E)$ is the effective area of the detector as a function of the energy and ${\cal M}(E',E)$ is the energy redistribution matrix, $E$ and $E'$ are respectively the photon energy\footnote{The values of $E'_{\rm min}$ and $E'_{\rm max}$ can differ from the values of $E_{\rm min}$ and $E_{\rm max}$.} and the measured value of photon energy.\\

We define the flux to count-rate conversion factor as $CR(z,Z,n_{\rm H},T_{500}) = S_{500}/L_{500}$.
However, clusters can be located at any position on the sky, and thus we have to convolve the $CR$ factor by the distribution of $n_{\rm H}$ on the sky. 
\begin{equation}
\overline{CR}(z,Z,T_{500}) = \int {\rm d}n_{\rm H} \, {\cal P}(n_{\rm H}) \,CR(z,Z,n_{\rm H},T_{500}),
\end{equation}
where ${\cal P}(n_{\rm H})$ is the probability to have a column density of hydrogen $n_{\rm H}$ on the line-of-sight. %This distribution is highly dependent on the sky fraction used for the analysis.  

\subsection{The tSZ-Xray cross power spectra}
\label{secth}
Decomposing both tSZ Compton parameter map and X-ray count-rate map we define
\begin{align}
y(\vec{n}) = \sum_{\ell m} y_{\ell m} Y_{\ell m} (\vec{n}),\\
x(\vec{n}) = \sum_{\ell m} x_{\ell m} Y_{\ell m} (\vec{n}).
\end{align}
Thus, the power spectra of both tSZ effect and X-ray can be written as
\begin{align}
C^{yy}_\ell = \frac{1}{2\ell +1} \sum_{m} y_{\ell m} y^{*}_{\ell m}\\
C^{xx}_\ell = \frac{1}{2\ell +1} \sum_{m} x_{\ell m} x^{*}_{\ell m},
\end{align}
the angular cross power spectrum of tSZ effect and X-ray count-rate map reads
\begin{equation}
C^{yx}_\ell = \frac{1}{2\ell +1} \sum_{m} \frac{1}{2}\left(y_{\ell m} x^{*}_{\ell m} + y^{*}_{\ell m} x_{\ell m}\right)
\end{equation}
To model this cross-correlation, or the auto correlation power spectra, we assume the following general expression
\begin{equation}
C_{\ell} = C^{{\rm 1h}}_\ell + C^{{\rm 2h}}_\ell + C^{{\rm diff}}_\ell,
\end{equation}
where $C^{{\rm 1h}}_\ell$ is the Poissonian contribution, $C^{{\rm 2h}}_\ell$ is the 2-halo term that account for correlation in the spatial distribution of clusters over the sky and $C^{{\rm diff}}_\ell$ is produced by the warm-hot intergalactic medium (WHIM). In the following, considering the low density and the low temperature of the WHIM, we assume $C^{{\rm diff}}_\ell << C^{{\rm 1h}}_\ell + C^{{\rm 2h}}_\ell$, and thus we neglect his contribution to the total power spectrum.\\

The Poissonian term can be computed by assuming the Fourier transform of tSZ and X-ray projected profiles weighted by the mass function, presented in Sect.~\ref{secmf}, and the fluxes for tSZ effect and X-ray count-rate \citep[see e.g,][for a derivation of the tSZ angular power spectrum]{kom02}.
{\small
\begin{equation}
C_{\ell}^{yx,{\rm 1h}} = 4 \pi \int {\rm d}z \frac{{\rm d}V}{{\rm d}z {\rm d}\Omega}\int{\rm d}M \frac{{\rm d^2N}}{{\rm d}M {\rm d}V} (1+\rho_{YL} \sigma_{{\rm log}\, Y}\sigma_{{\rm log}\, L}){Y}_{500} {S}_{500} y_{\ell} x_{\ell},
\end{equation}
}
where ${S}_{500} = \overline{CR}(z,Z,T_{500})  {L}_{500}$, $\overline{CR}(z,Z,T_{500})$ the flux to count-rate conversion factor described in Sect.~\ref{secbrem}, $\frac{{\rm d^2N}}{{\rm d}M {\rm d}V}$ the clusters mass function described in Sect.~\ref{secmf}, and $\frac{{\rm d}V}{{\rm d}z {\rm d}\Omega}$ the element of comoving volume. The term $(1+\rho_{YL} \sigma_{{\rm log}\, Y}\sigma_{{\rm log}\, L})$ accounts for extra-power produced by the scatter in the scaling relations described at Sect.~\ref{secscal}.\\
The Fourier transform of the 3-D profile projected across the line-of-sight on the sphere reads,
\begin{equation}
p_{\ell} = \frac{4 \pi r_{\rm o}}{l^2_{\rm s}} \int_0^{\infty} {\rm d}r_{\rm s} \, r_{\rm s}^2 p(r_{\rm s}) \frac{{\rm sin}(\ell r_{\rm s} / \ell_{\rm s})}{\ell r_{\rm s} / \ell_{\rm s}},
\end{equation}
where $p(r_{\rm s})$ is either the tSZ 3-D profile or the X-ray count-rate 3-D profile, $r_{\rm s} = r/r_{\rm o}$, $\ell_{{\rm s}} = D_{\rm A}(z)/r_{\rm o}$, $r_{\rm o}$ is the scale radius of the profile.\\

The contribution of the 2-halo term corresponds to large scale fluctuations in the matter power spectrum, that induce correlation in the cluster distribution over the sky.
It can be computed as \citep[see e.g,][and references therein]{tab11}
\begin{align}
C_{\ell}^{yx,{\rm 2h}} = 4 \pi \int {\rm d}z \frac{{\rm d}V}{{\rm d}z{\rm d}\Omega}&\left(\int{\rm d}M \frac{{\rm d^2N}}{{\rm d}M {\rm d}V} {Y}_{500} y_{\ell} b_{\rm lin}(M,z)\right)\\ \nonumber
&\times \left(\int{\rm d}M \frac{{\rm d^2N}}{{\rm d}M {\rm d}V} {S}_{500} x_{\ell} b_{\rm lin}(M,z)\right) P(k,z)
\end{align}
with $P(k,z)$ the matter power-spectrum computed using $\textsc{CLASS}$ \citep{les11} and $b_{\rm lin}(M,z)$ the time dependent linear bias factor that relates $P(k,z)$ to the power spectrum of the cluster distribution over the sky. 
Following \citet{mo96,kom99}, we adopt $b_{\rm lin}(M,z)=1+(\nu^2(M,z)-1)/\delta_c(z)$, whith $\nu(M,z) = \delta_c(z)/\left[D_g(z) \sigma(M)\right]$, $D_{\rm g}(z)$ is the linear growth factor and $\delta_c(z)$ is the over-density threshold for spherical collapse.\\

\subsection{Mass function}
\label{secmf}
Our computation of the tSZ-X correlation assumes the mass function calibrated on numerical simulation from \citet{tin08},
\begin{equation}
\frac{{\rm d} N}{{\rm d}M_{500}{\rm d} V} = f(\sigma)\frac{\rho_{\rm m}(z=0)}{M_{500}}\frac{{\rm d ln} \sigma^{-1} }{{\rm d}M_{500}},
\end{equation}
with
\begin{equation}
f(\sigma) = A_0\left[ 1+ \left( \frac{\sigma}{A_1}\right)^{A_2}\right] {\rm exp}\left(-\frac{A_3}{\sigma^2} \right),
\end{equation}
and $\rho_{\rm m}(z = 0)$ the mean matter density today. The coefficients $A_0$, $A_1$, $A_2$ and $A_3$ are given in \citet{tin08} for various over densities, $\Delta_{\rm mean}$, with respect to the redshift dependent mean cosmic density. These coefficients are interpolated to match $\Delta_{\rm critical}$ defined with respect to the critical density. The relation between $\Delta_{\rm critical}$ and $\Delta_{\rm mean}$ is given by $\Delta_{\rm mean} = \Delta_{\rm critical}/\Omega_{\rm m}(z)$, with $\Omega_{\rm m}(z)$ the matter density parameter at redshift $z$.
The standard deviation of the density perturbation in a sphere or radius $R$, $\sigma$, is computed in linear perturbation theory.

\subsection{tSZ and X-ray fluxes}
\label{secscal}
A key step in the modeling of the cross correlation between tSZ and X-ray is to relate the mass, $M_{500}$, and the redshift, $z$, of a given cluster to tSZ flux, $Y_{500}$, and X-ray luminosity $L_{500}$ in the $\left[E_{\rm min}-E_{\rm max}\right]$~keV energy band\footnote{By convention, the energy range [0.1-2.4] keV is used for the definition of the $M_{500}-L_{500}$ scaling relation.} in the rest frame of the cluster. 
The cross correlation between tSZ effect and X-ray emission is thus highly dependent on the $M_{500}-Y_{500}$ and the $M_{500}-L_{500}$ relations in terms of normalization and slope.
Consequently, we need to use the relations derived from a representative sample of galaxy clusters, with a careful propagation of statistical and systematic uncertainties. 
We stress that for power spectrum analysis the intrinsic scatter of such scaling laws has to be considered, as it will produce extra-power that has to be accounted for in order to avoid biases.\\
We used the $M_{500}-Y_{500}$ scaling laws presented in \citet{PlanckSZC}, 
\begin{equation}
E^{-\beta_{\rm sz}}(z) \left[ \frac{D^2_{ang}(z) {Y}_{500}}{10^{-4}\,{\rm Mpc}^2} \right] = Y_\star \left[ \frac{h}{0.7} \right]^{-2+\alpha_{\rm sz}} \left[ \frac{(1-b) M_{500}}{6 \times 10^{14}\,{\rm M_{\odot}}}\right]^{\alpha_{\rm sz}},
\label{szlaw}
\end{equation}
with $E(z) = \Omega_{\rm m}(1+z)^3 + \Omega_{\Lambda}$. The coefficients $Y_\star$, $\alpha_{\rm sz}$ and $\beta_{\rm sz}$, from \citet{PlanckSZC}, are given in Table~\ref{tabscal}. We used $b=0.2$ for the bias between X-ray estimated mass and effective mass of the clusters.
To model the $L_{500}-M_{500}$ relation we used the relation derived by \citet{arn10} from the \textsc{REXCESS} sample \citep{boh07}.
\begin{equation}
\left[h E(z)\right]^{-\beta_{\rm x}}\left(\frac{{L}_{500}}{{10^{44}}\,{\rm erg.s^{-1}}}\right) = L_\star \left[ \frac{M_{{\rm x},500}}{3 \times 10^{14}\,{\rm M_{\odot}}}\right]^{\alpha_{\rm x}},
\end{equation}
where $M_{{\rm x},500}$ is the cluster mass estimated from X-ray observations. It is related to the true mass $M_{500}$ through $M_{{\rm x},500} = (1-b) M_{500}$
the coefficients $L_\star$, $\alpha_{\rm x}$ and $\beta_{\rm x}$ are given in Table.~\ref{tabscal}.\\
The two relations, $M_{500}-Y_{500}$ and $M_{500}-L_{500}$, have intrinsic scatters, $\sigma_{{\rm log}\, Y} = 0.075$ and $\sigma_{{\rm log}\, L} = 0.183$, respectively. These scatters will contribute to the total power measured on the sky. Indeed, the quantity $<{Y^2}_{500}>$ is equal to $(1+\sigma^2_{{\rm log}\, Y}){Y}^2_{500}$ and $<{L^2}_{500}>$ is equal to $(1+\sigma^2_{{\rm log}\, L}){L}^2_{500}$.  
The cross-correlation power spectrum of tSZ and X-ray can be affected by the same effect. However, for a cross correlation this scatter bias is dependent of the correlation between the scatters of the $M_{500}-Y_{500}$ and the $M_{500}-L_{500}$ scaling relations. This question is addressed in Sect.~\ref{sectszx}.\\

We also need to have an estimate of the cluster temperature, ${T}_{500}$. In this work, we used the scaling law from \citet{planckSL}
\begin{equation}
E(z)^{-\beta_{\rm T}} {Y}_{500} = T_{\star} \left[ \frac{{T}_{500}}{6\,{\rm keV}} \right]^{\alpha_{\rm T}},
\end{equation}
the coefficients $T_\star$, $\alpha_{\rm T}$ and $\beta_{\rm T}$ are given in Table~\ref{tabscal}.
This relation also presents an intrinsic scatter $\sigma_{{\rm log}\, T} = 0.14 \pm 0.02$. We verified that this scatter has no significant impact on the tSZ-X cross power-spectrum amplitude with respect to the scatter from $M_{500}-Y_{500}$ and $M_{500}-L_{500}$ relations.

\begin{table*}
\center
\caption{Scaling-law parameters and error budget for both $Y_{500}-M_{500}$ \citep{PlanckSZC}, $L_{500}-M_{500}$ \citep{arn10} and $Y_{500}-T_{500}$ \citep{PlanckSZC} relations}
\begin{tabular}{|cc|cc|cc|}
\hline
\multicolumn{2}{|c|}{$M_{500}-Y_{500}$} & \multicolumn{2}{c}{$M_{500}-L_{500}$} & \multicolumn{2}{|c|}{$M_{500}-T_{500}$} \\
\hline
${\rm log}\,Y_\star$ & -0.19 $\pm$ 0.02 & ${\rm log}\, L_\star$ & 0.724 $\pm$ 0.032 & ${\rm log}\, T_\star$ & -4.27 $\pm$ 0.02   \\
$\alpha_{\rm sz}$ & 1.79 $\pm$ 0.08 & $\alpha_{\rm x}$ & 1.64 $\pm$ 0.12 & $\alpha_{\rm T}$ & 2.85 $\pm$ 0.18  \\
$\beta_{\rm sz}$ & 0.66 $\pm$ 0.50 & $\beta_{\rm x}$ & 7/3 &  $\beta_{\rm T}$ & 1 \\
$\sigma_{{\rm log}\, Y}$ & 0.075 $\pm$ 0.010 & $\sigma_{{\rm log}\, L}$ & 0.183 $\pm$ 0.032 & $\sigma_{{\rm log}\, T}$ & 0.14 $\pm$ 0.02 \\
\hline
\end{tabular}
\label{tabscal}
\end{table*}

\subsection{Pressure and density profiles}

The tSZ effect is directly proportional to the pressure integrated across the line-of-sight. In this work, we model the galaxy cluster pressure profile by a Generalized Navarro Frenk and White \citep[GNFW,][]{nav97,nag07} profile of the form
\begin{equation}
{\mathbb P}(r) = \frac{P_0}{\left(c_{500} r\right)^\gamma \left[1 + (c_{500} r)^\alpha \right]^{(\beta-\gamma)/\alpha}}.
\end{equation}
For the parameters $c_{500}$, $\alpha$, $\beta$, and $\gamma$, we used the best fitting values from \citet{arn10} presented in Table.~\ref{tabscal}. The absolute normalization of the profile $P_0$ is set assuming the scaling laws $Y_{500}-M_{500}$ presented in Sect.~\ref{secscal}.\\
To model the X-ray emission, we need both the density, $n_e(x)$, and the temperature, ${T_e}(x)$, profiles. Thus, we assume a polytropic profile \citep[see e.g;][]{kom01}, ${\mathbb P}(r) = n_e(r)T_e(r)$, with
$n_e(r) \propto T_e(r)^\delta$ where $\delta$ the polytropic index fixed at 1.5.
For the X-ray flux to count-rate conversion factor, we only assume an averaged temperature $T_{500}$, then the X-ray flux is directly proportional to $n^2_e(x)$. The overall normalization of the X-ray emission profile is deduced from the scaling law $L_{500}-M_{500}$ presented in Sect.~\ref{secscal}.

\section{Results}
\label{secres}
\subsection{The tSZ-X power spectrum}
\label{sectszx1}
In Fig.~\ref{crossxsz}, we present the angular cross power spectrum between tSZ  and X-rays (assuming 0.5-2.0 keV energy band for the ROSAT experiment). 
The power spectrum is predicted for our fiducial cosmological model.
We observe that the tSZ-X power spectrum is dominated by the 1-halo term throughout the entire range of multiplole, from $\ell = 0$ to $\ell = 10\,000$.\\
For low multipoles ($\ell$ < $1\,000$) the tSZ-X power spectrum follows a power law $C^{yx}_\ell \propto \ell^{-1.1}$.

The correlation factor between tSZ and X-ray surveys, $\rho_{yx} = \frac{C_\ell^{yx}}{\sqrt{C_\ell^{yy}C_\ell^{xx}}}$, as a function of $\ell$ presents a slow variation from 0.8 at low $\ell$ to 0.6 at high $\ell$. The smaller correlation factor at small scales highlight the difference of slopes in tSZ and X-ray profiles in the core of the cluster, as the tSZ profile decreases with $T_{\rm e} n_{\rm e}$ and the X-ray profile with $n_{\rm e}^2$. \\

\subsection{The scaling relation scatter bias}
\label{sectszx}
The amplitude of the 1-halo term of tSZ-X power spectrum is sensitive to the scatter of scaling relations, which produce an excess of power. 
The 2-halo term is not affected by the scatter.\\
For tSZ auto-correlation power spectrum the scatter produces a negligible bias of $0.5$\%. For the X-ray power spectrum the effect reach $3.3$\%.\\
The bias on tSZ-X cannot be estimated as easily. Indeed the correlation between $M_{500}-Y_{500}$ and $M_{500}-L_{500}$ scatters has to be known.
The quantity $<{Y_{500} \, L_{500}}>$ is equal to $(1+\rho_{YL} \sigma_{{\rm log}\, Y}\sigma_{{\rm log}\, L}){Y}_{500}\, {L}_{500}$, where $\rho_{YL}$ is the correlation between the $M_{500}-Y_{500}$ and $M_{500}-L_{500}$ scatters.
Consequently, the bias is null for a 0 correlation and maximal for a full correlation.\\
Using the $Y_{500}-L_{500}$ scatter, $\sigma_{{\rm log}\, YL} = 0.14 \pm 0.02$ \citep{planckSL}, it is possible to estimate $\rho_{YL}$ through
\begin{equation}
\rho_{YL} = \frac{\sigma_{{\rm log}\, Y}^2 \left( \frac{\alpha_{\rm SZ}}{\alpha_{\rm X}}\right)^2 +  \sigma_{{\rm log}\, L}^2 - \sigma_{{\rm log}\, YL}^2}{2  \frac{\alpha_{\rm SZ}}{\alpha_{\rm X}}  \sigma_{{\rm log}\, Y} \sigma_{{\rm log}\, L} } = 0.9 \pm 0.3.
\end{equation} 
This finding is consistent with an almost full correlation between $M_{500}-Y_{500}$ and $M_{500}-L_{500}$ scatters. This value for $\rho_{YL}$ leads to a bias of $1.4$\% for the amplitude of the tSZ-X power spectrum.

\subsection{Redshift and mass distribution of the tSZ-Xray cross correlation}
\label{secdis}

\begin{figure*}[!th]
\begin{center}
\includegraphics[scale=0.4]{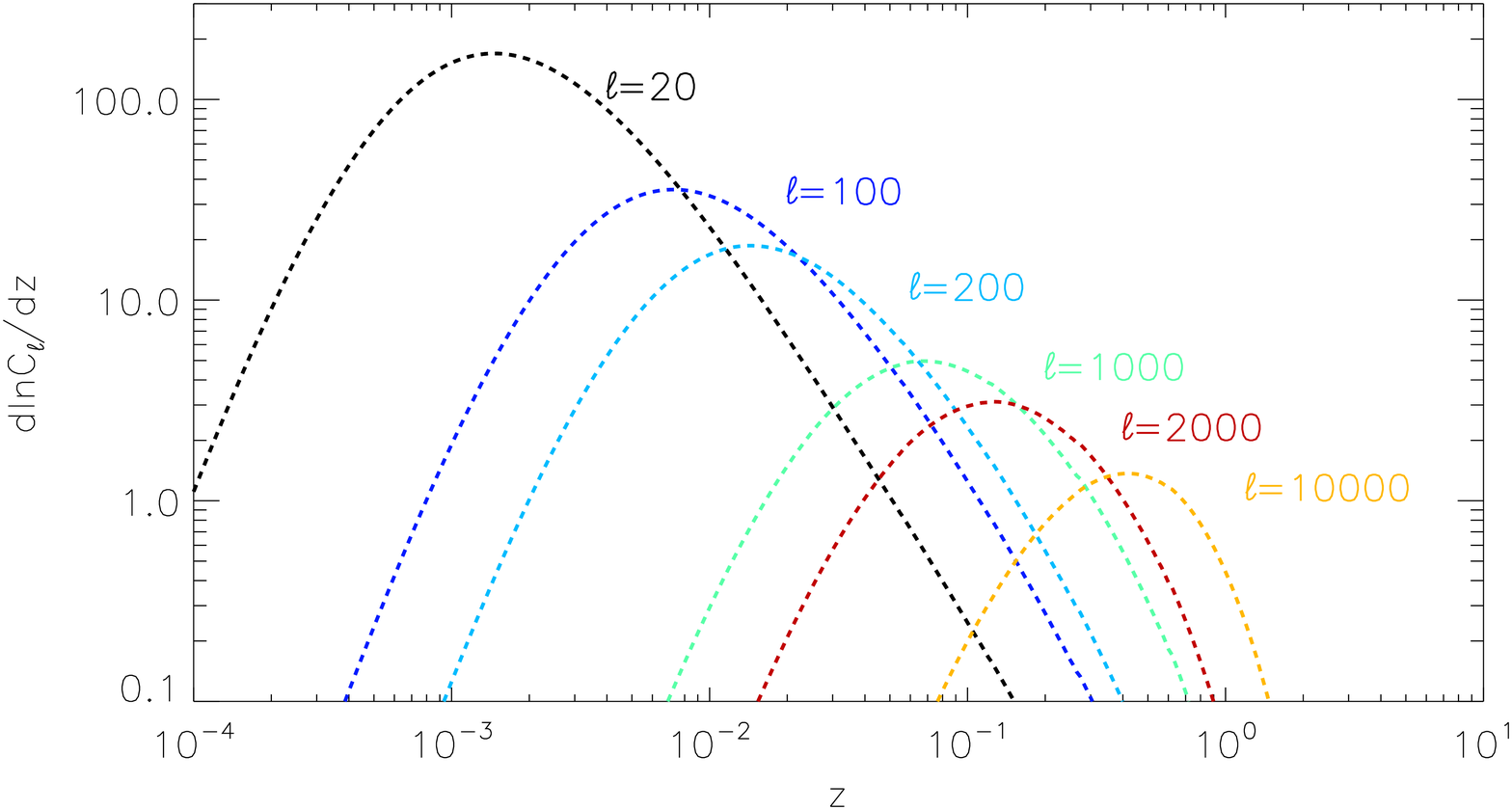}
\caption{Redshift distribution of the contribution to the total tSZ/Xray cross-correlation power for several values of $\ell$. In black for $\ell=20$, in dark blue for $\ell = 100$, in light blue for $\ell = 200$, in green for $\ell = 1000$, in orange for $\ell = 2000$ and in red for $\ell = 10\,000$.}
\label{reddis}
\end{center}
\end{figure*}

\begin{figure*}[!th]
\begin{center}
\includegraphics[scale=0.4]{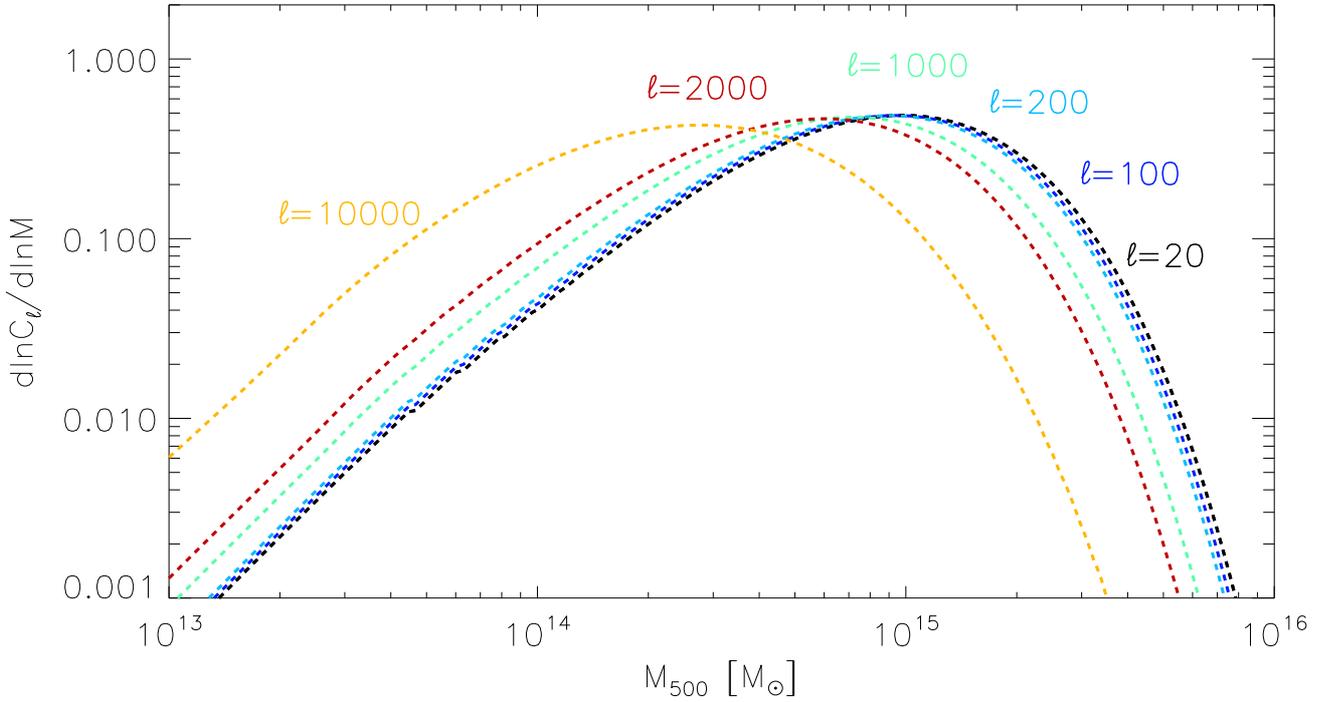}
\caption{Mass distribution of the contribution to the total tSZ/Xray cross-correlation power for several values of $\ell$. In black for $\ell=20$, in dark blue for $\ell = 100$, in light blue for $\ell = 200$, in green for $\ell = 1000$, in orange for $\ell = 2000$ and in red for $\ell = 10\,000$.}
\label{massdis}
\end{center}
\end{figure*}

In Fig.~\ref{reddis} and Fig.~\ref{massdis}, we present the distribution of the tSZ-Xray cross correlation power spectrum as function of the redshift and the mass for various values of $\ell$ from 20 to $10\,000$.  
We observe that the power below $\ell = 100$ is dominated by local object at redshifts below 0.2. Whereas at high multipole values, $\ell = 10 000$, we are sensitive to structures up to $z=1.5$. 
We observe that the small and large angular scales of the power spectrum are sampling distinct populations in terms of redshift.\\
Contrary to the redshift distribution, we observe that the mass dependency presents small variations for $\ell$ ranging from 20 to 2000. For these multipoles the power is dominated by objects with $M_{500}$ above $10^{14}$ M$_\odot$. 
Smaller mass objects only have significant contribution to the total power for very high multipoles value $\ell \simeq 10\,000$.\\

%%%%%%%%%%%%%

The mass-function predicts a number of objects above a given mass $M_{500}$ that drastically increases when $M_{500}$ decreases. Similarly, the comoving volume increases with increasing redshift (below $z \simeq 2$).
The 2-halo term presented in Sect.~\ref{secth} is proportional to the number of clusters, that significantly contributes to the total power, to the square, contrary to the 1-halo terms that is linearly related to this quantity.\\ 
This explains the relative small amplitude of the 2-halo term with respect to the 1-halo term for the tSZ-X power spectrum. Indeed, the low redshift depth and the high mass sensitivity of the tSZ-X power spectrum imply that the total power is dominated by a small number of objects.\\
Consequently, small values of $\alpha_{\rm x}$ or $\alpha_{\rm sz}$, promote the 2-halo term with respect to the 1-halo term. The 2-halo term becomes significant at low-$\ell$ for $\alpha_{\rm x} + \alpha_{\rm sz} < 3$. 
However, such values are excluded by existing constraints on the the $M_{500}-Y_{500}$ and the $M_{500}-L_{500}$ relations (see Table.~\ref{tabscal} for allowed uncertainty range and Fig.~\ref{crossxsz} for the impact on tSZ-X cross power spectrum of these uncertainties).

\subsection{tSZ-X cross correlation dependencies with modeling parameters}
\label{secdep}

\begin{figure}[!th]
\begin{center}
\includegraphics[scale=0.2]{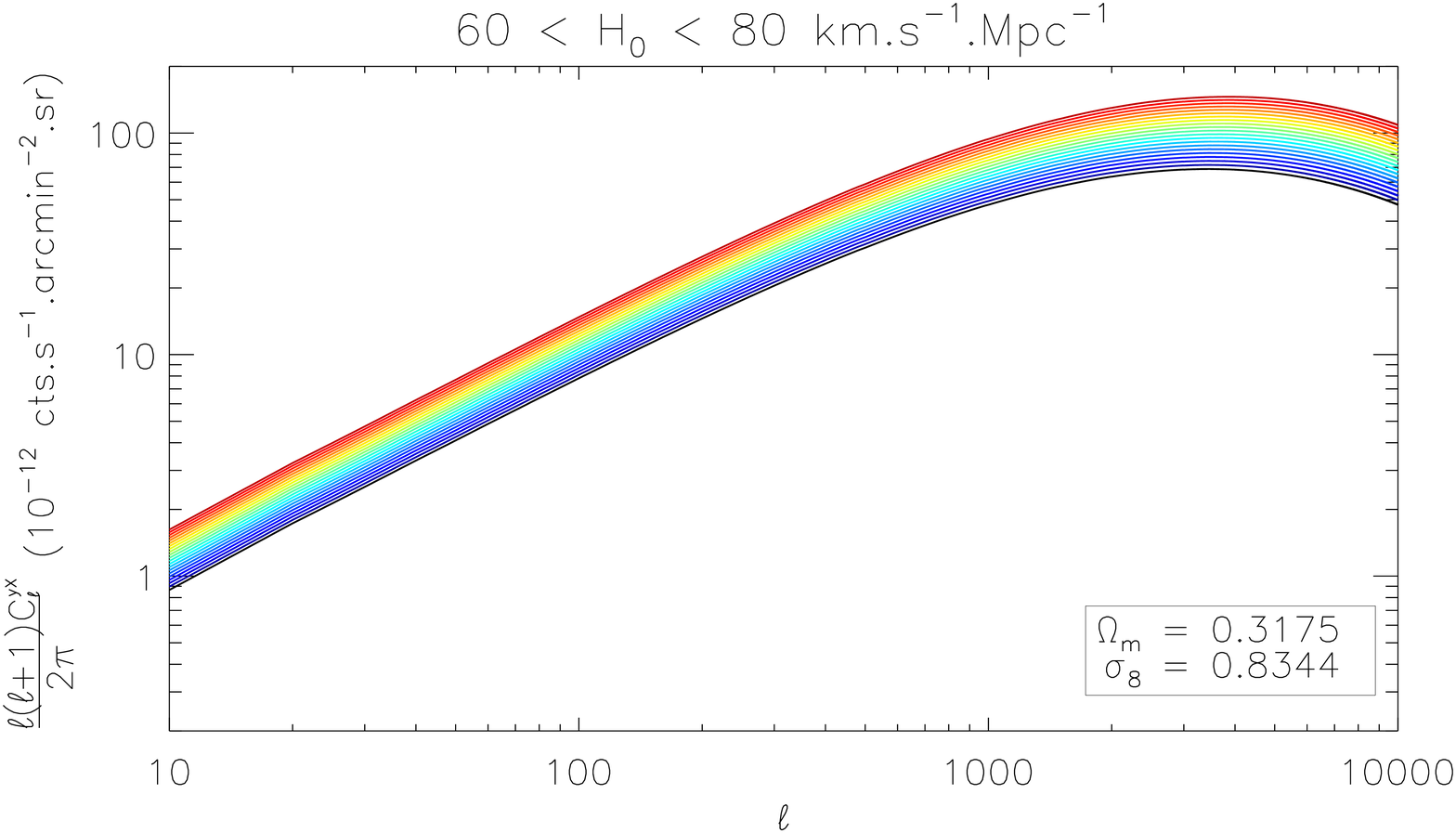}
\includegraphics[scale=0.2]{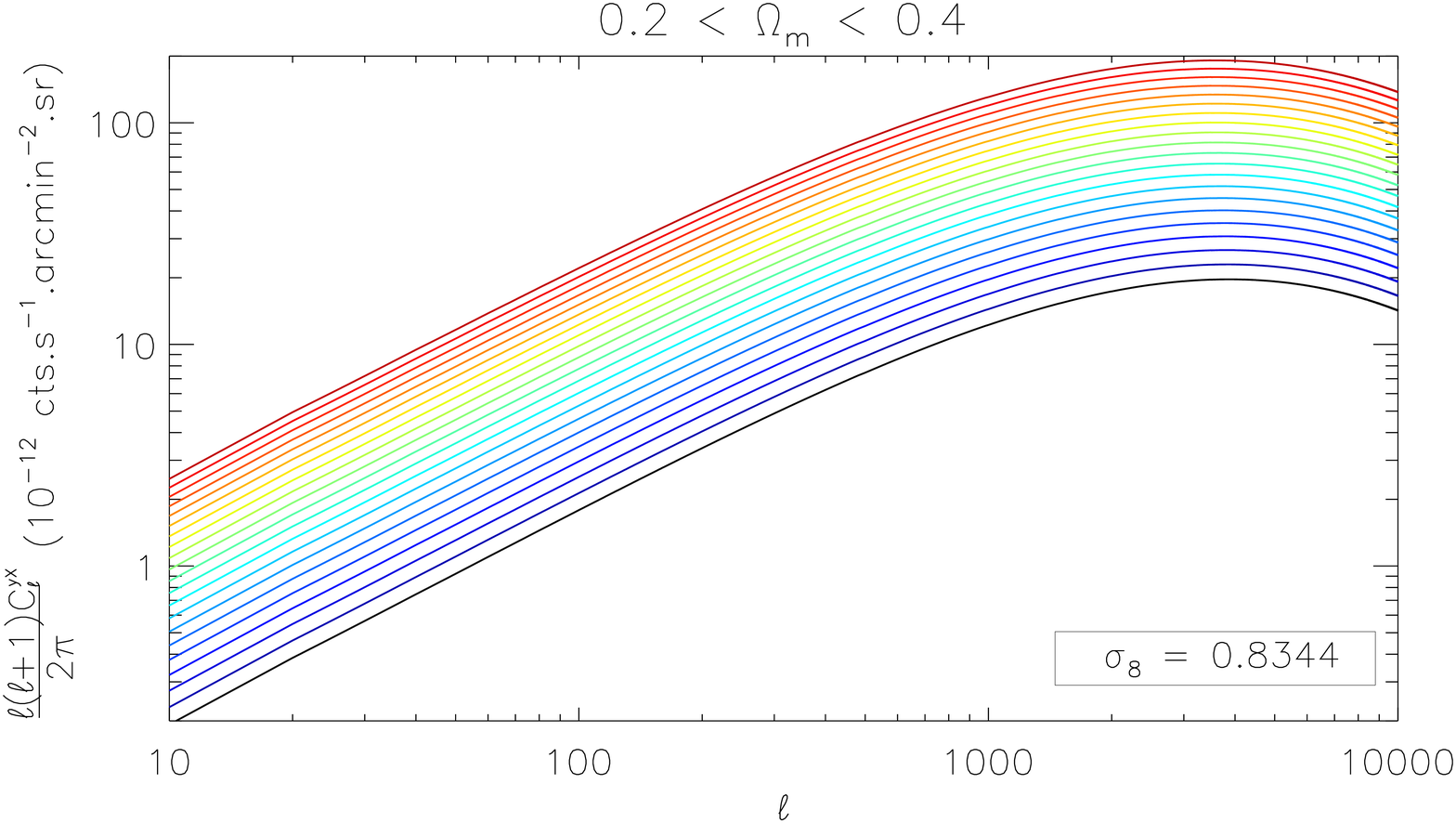}
\includegraphics[scale=0.2]{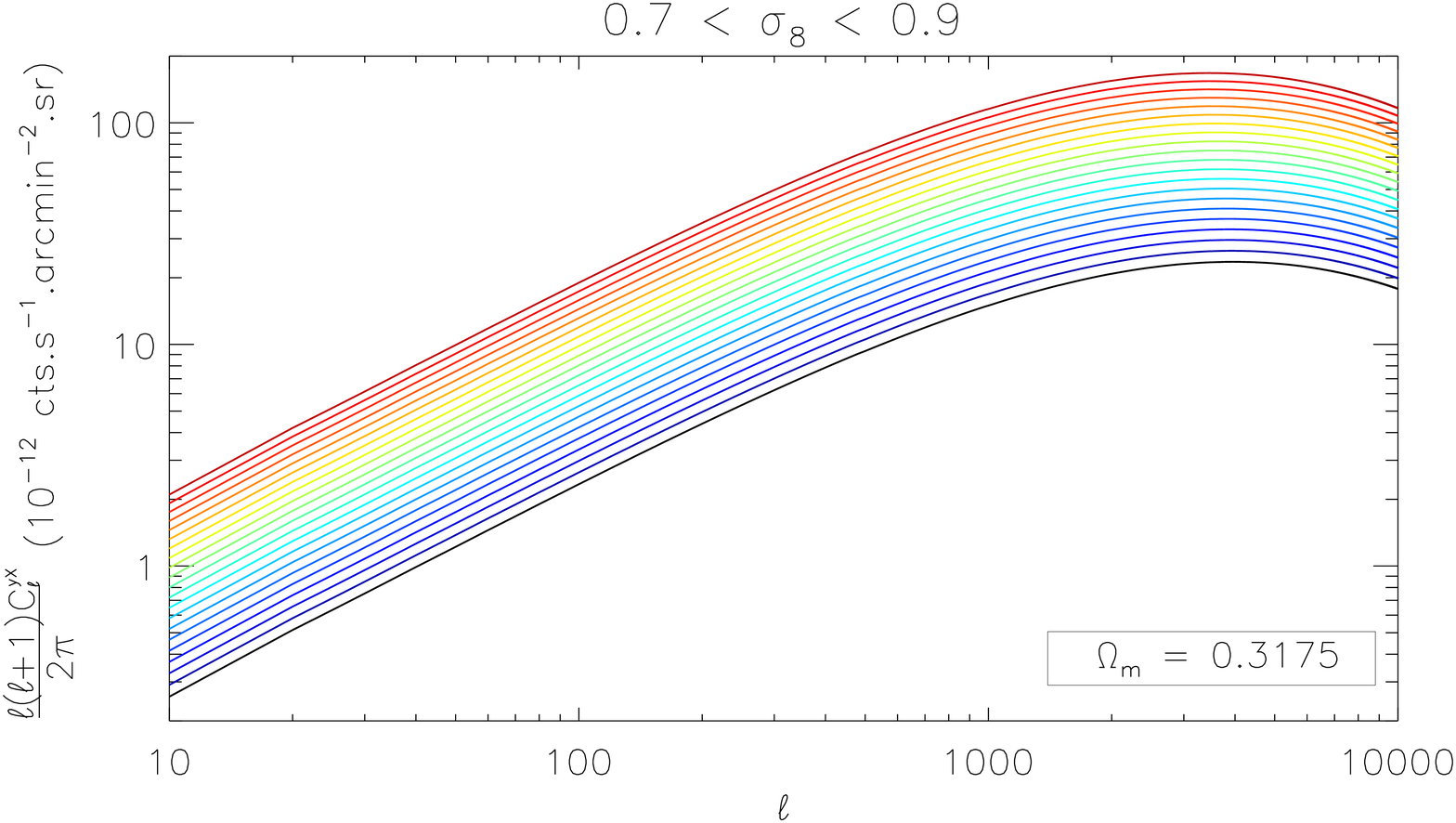}\\
\caption{Theoretical tSZ-Xray cross correlation power spectra, as a function of, from top to bottom, $H_0$, $\Omega_{\rm m}$ and $\sigma_8$. From blue to red for 21 values of $H_0$, $\Omega_{\rm m}$ and $\sigma_8$, starting at 60 km/s.Mpc, 0.2, and 0.7 with a step of 1 km/s/Mpc, 0.01 and 0.01 respectively.}
\label{parvar}
\end{center}
\end{figure}

Our modeling is affected both by cosmological and scaling law parameters.\\
First, we focus on cosmological parameters, with a particular attention to $H_0$, $\Omega_{\rm m}$ and $\sigma_8$.
In Fig.~\ref{parvar}, we present the variation of the tSZ-X cross spectrum as a function of $H_0$ from 60 to 80 km/s/Mpc with a step of 1 km/s/Mpc, $\Omega_{\rm m}$ from 0.2 to 0.4 with a step of 0.01 and $\sigma_8$ from 0.7 to 0.9 with a step of 0.01. 
In the most general case, those variations depend of the multipole $\ell$, as presented in Figs.~\ref{reddis}~and~\ref{massdis} each multipole is sensitive to different regions of the mass function and thus present different sensitivity to the cosmological parameters. 
\begin{equation}
A_{\rm cl} \propto \left( \frac{H_0}{67} \right)^{\alpha_H(\ell)} \left( \frac{\Omega_{\rm m}}{0.32} \right)^{\alpha_\Omega(\ell)}  \left( \frac{\sigma_8}{0.83} \right)^{\alpha_\sigma(\ell)},
\label{cosmoeq}
\end{equation}
where $A_{\rm cl}$ is the amplitude of the tSZ-X power spectrum.
However as shown in Fig.~\ref{parvar}, we do not observe a significant distortion of the shape of the cross-correlation with a variation of the cosmological parameters.\\
Similar expressions can be used for tSZ and X-ray auto-correlation spectra.
In table.~\ref{powlaw} we present the values of $\alpha_H(\ell)$, $\alpha_{\Omega}(\ell)$, and $\alpha_{\sigma}(\ell)$ for each spectra, tSZ-X, tSZ-auto, X-auto. \\

\begin{figure}[!th]
\begin{center}
\includegraphics[scale=0.2]{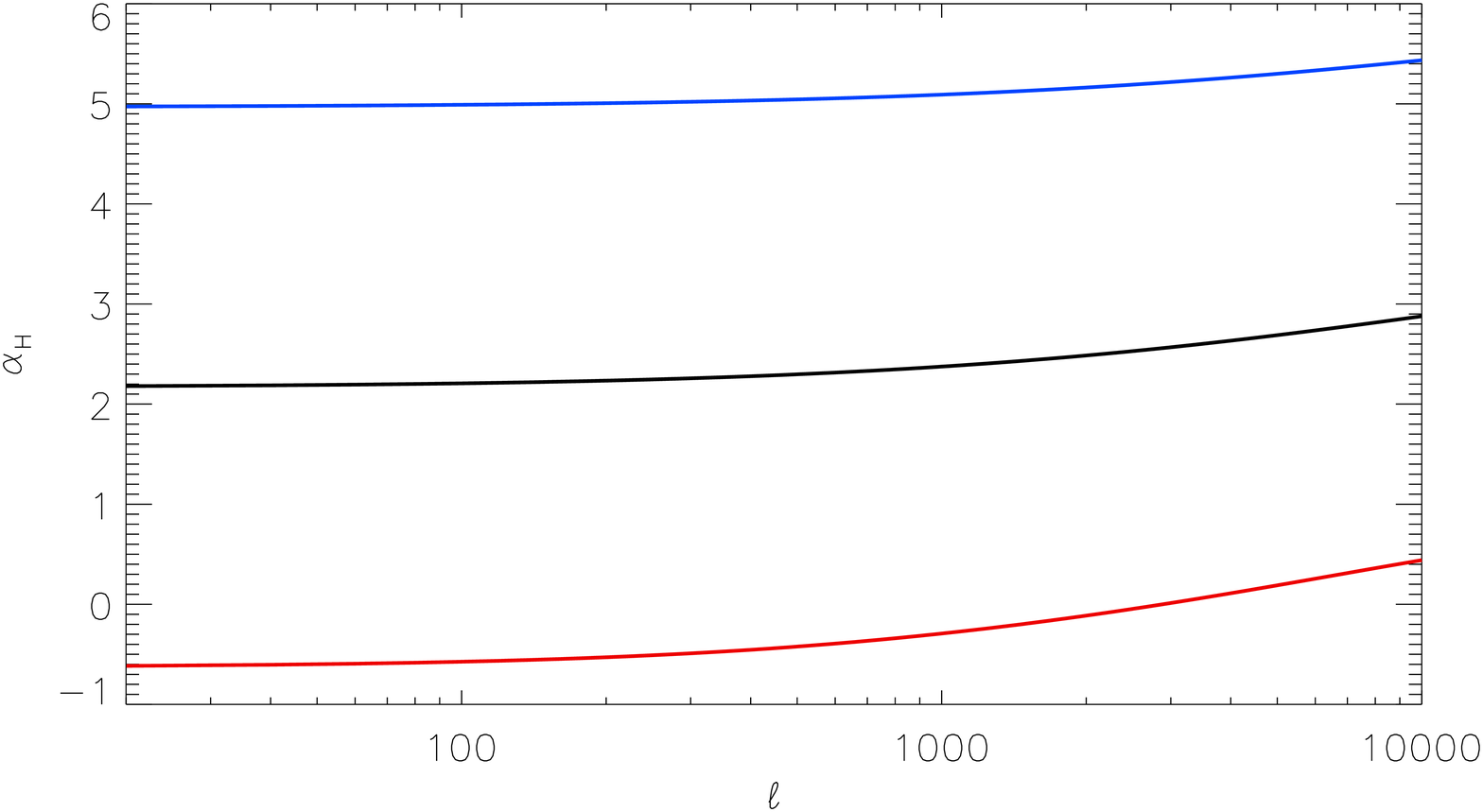}
\includegraphics[scale=0.2]{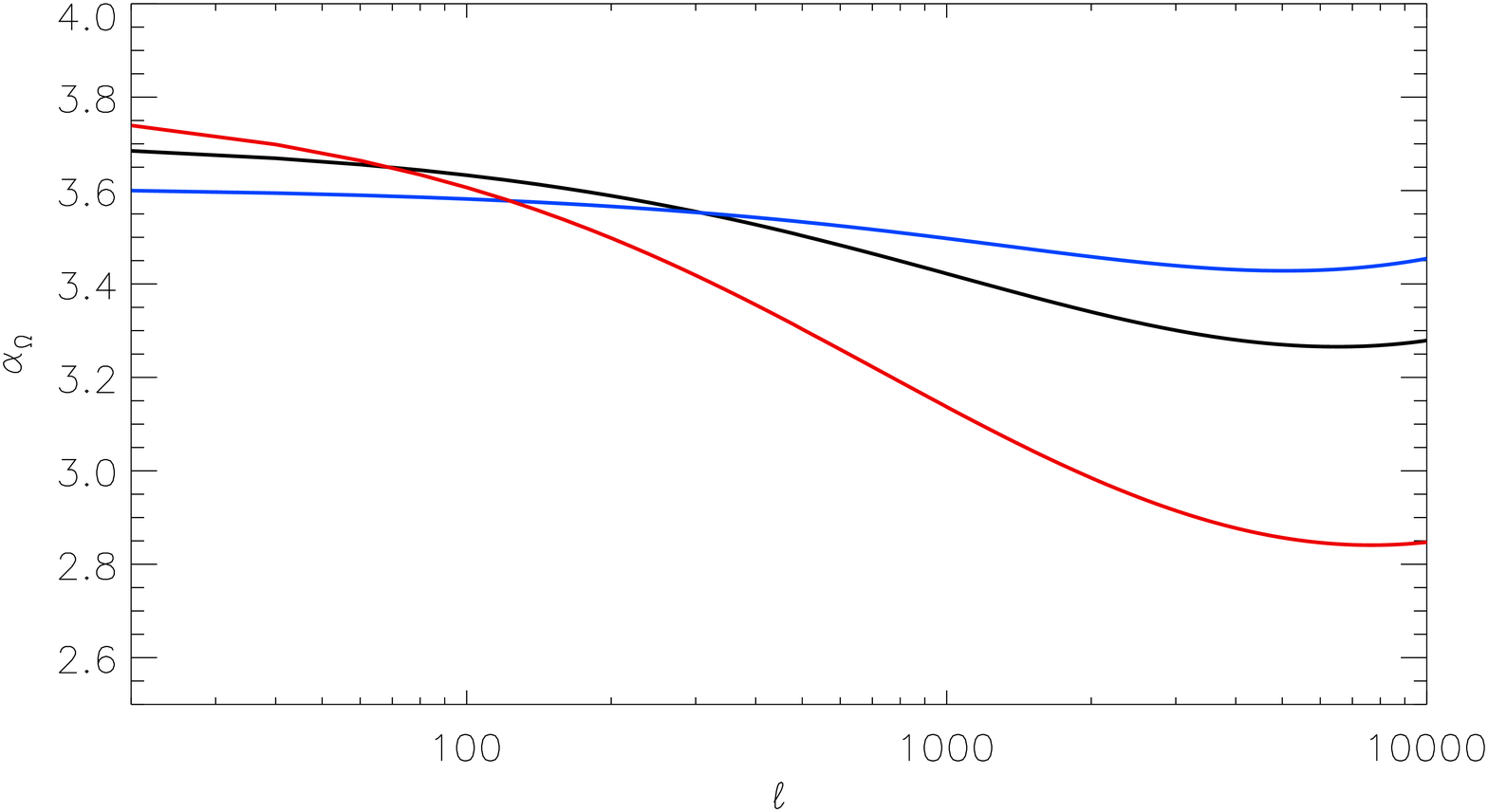}
\includegraphics[scale=0.2]{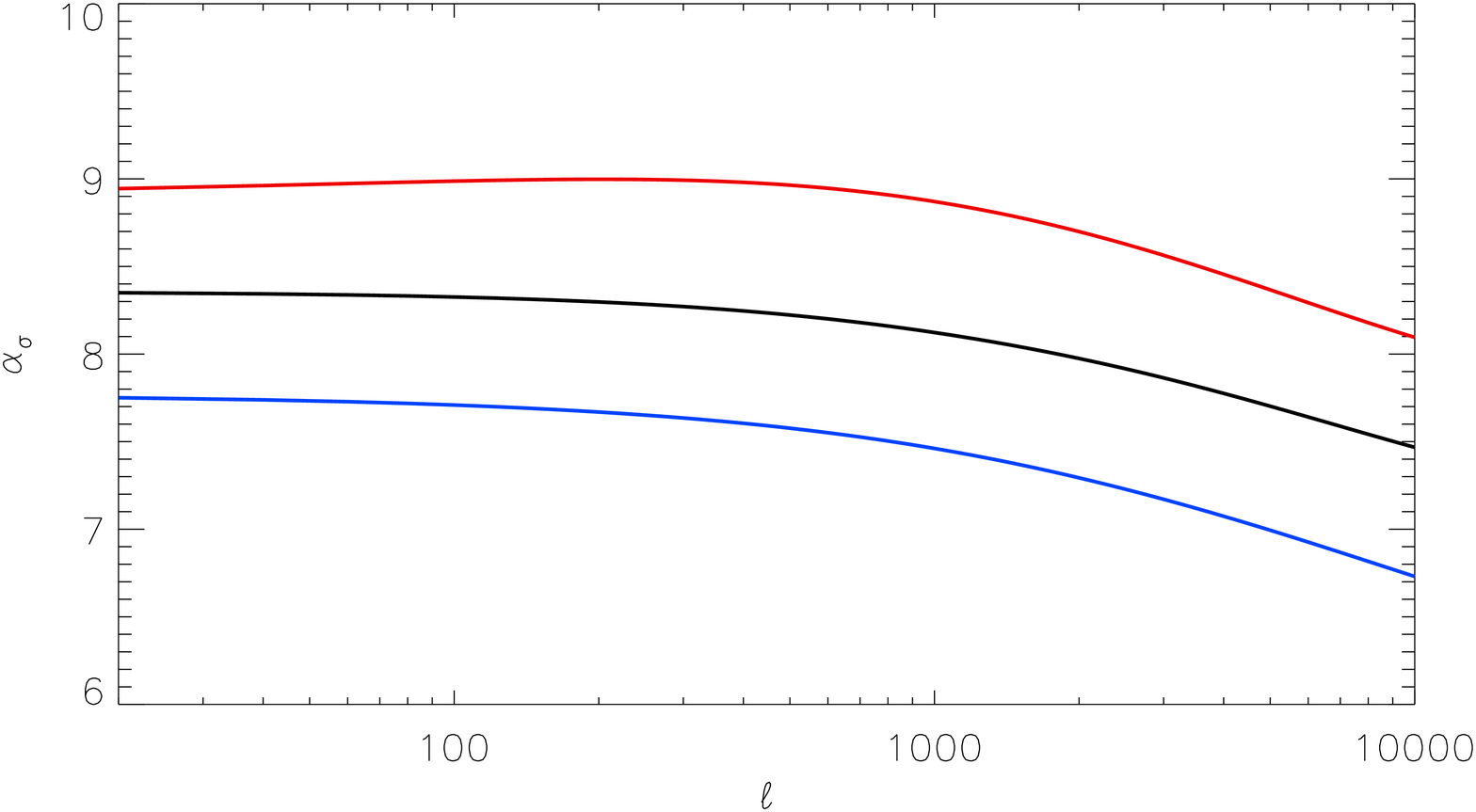}
\caption{Power law indexes variation as a function of $\ell$ for in black the X-tSZ cross-correlation, in red the tSZ auto-correlation and in blue the X-ray auto-correlation. From top to bottom for $\alpha_H(\ell)$, $\alpha_\Omega(\ell)$ and $\alpha_\sigma(\ell)$.}
\label{pslaw}
\end{center}
\end{figure}

In Fig.~\ref{pslaw}, we present the power law index variations with respect to $\ell$. For the tSZ auto-correlation we observe $\alpha_H\simeq 0$, however both X-ray auto-spectrum and X-tSZ cross spectrum present significant variations with $H_0$, with $\alpha^{X}_H \simeq 5$ and $\alpha^{XSZ}_H \simeq 2.5$. These dependencies variation are produced by the $L_{500} - M_{500}$ scaling relation. 
We also observe that $\alpha_\Omega$ ranges from 2.8 to 3.8 and $\alpha_\sigma$ from 7 to 9.
These variation of spectral indexes with respect to $\ell$ are small. 

In Table.~\ref{powlaw}, we provide fitting formula for each power law index with respect to $\ell$ using the following parametric formula.
\begin{equation}
\alpha = p_1 + p_2 \left( \ell + p_3 \right)^{p_4}.
\label{parfor}
\end{equation}
Values for $p_1$, $p_2$, $p_3$, $p_4$ are provided in Table.~\ref{powlaw}.
We note that the power law indexes are also function of the cosmological parameters. Consequently, we stress the fact that the formula given above have been estimated for cosmological parameters ($H_0$, $\Omega_{\rm m}$,$\sigma_8$) = (67,0.32,0.83), and thus can only be considered accurate for 10\% variations around these parameters.
\begin{table}
\center
\caption{Power law indexes of the tSZ-X, tSZ-auto, and X-auto power spectra for variations with from top to bottom $H_0$, $\Omega_{\rm m}$, and $\sigma_8$ as described in Eq.~\ref{cosmoeq} and Eq.~\ref{parfor}.}
\begin{tabular}{|c|ccccc|}
\hline
 & $\overline{\alpha_H}$ & $p_1$ & $p_2$ & $p_3$ & $p_4$ \\
 \hline
tSZ-X & 2.38 & -0.025 & 1.027 & 967 & 0.1114 \\
\hline
tSZ-auto & -0.29 & -3.60 & 1.50 & 597 & 0.1071 \\
\hline
X-auto & 5.09 & 0 & 3.475 & 1998 & -0.04747 \\
\hline
\hline
 & $\overline{\alpha_\Omega}$ & $p_1$ & $p_2$ & $p_3$ & $p_4$ \\
 \hline
tSZ-X & 3.42 & 3.24 & 1.66$\times 10^5$ & 1426 & -1.767 \\
\hline
tSZ-auto & 3.14 & 2.836 & 7.99$\times 10^7$ & 1680 & -2.463 \\
\hline
X-auto & 3.50 & 3.42 & 1.66$\times 10^5$ & 1704 & -1.844 \\
\hline
\hline
 & $\overline{\alpha_\sigma}$ & $p_1$ & $p_2$ & $p_3$ & $p_4$ \\
 \hline
tSZ-X & 8.12 & 0 & 12.55 & 1498 & -0.0557 \\
\hline
tSZ-auto & 8.87 & 0 & 15.06 & 2017 & -0.0660 \\
\hline
X-auto & 7.46 & 0 & 12.12 & 1190 & -0.0620 \\
\hline
\end{tabular}
\label{powlaw}
\end{table}

\begin{figure}[!th]
\begin{center}
\includegraphics[scale=0.2]{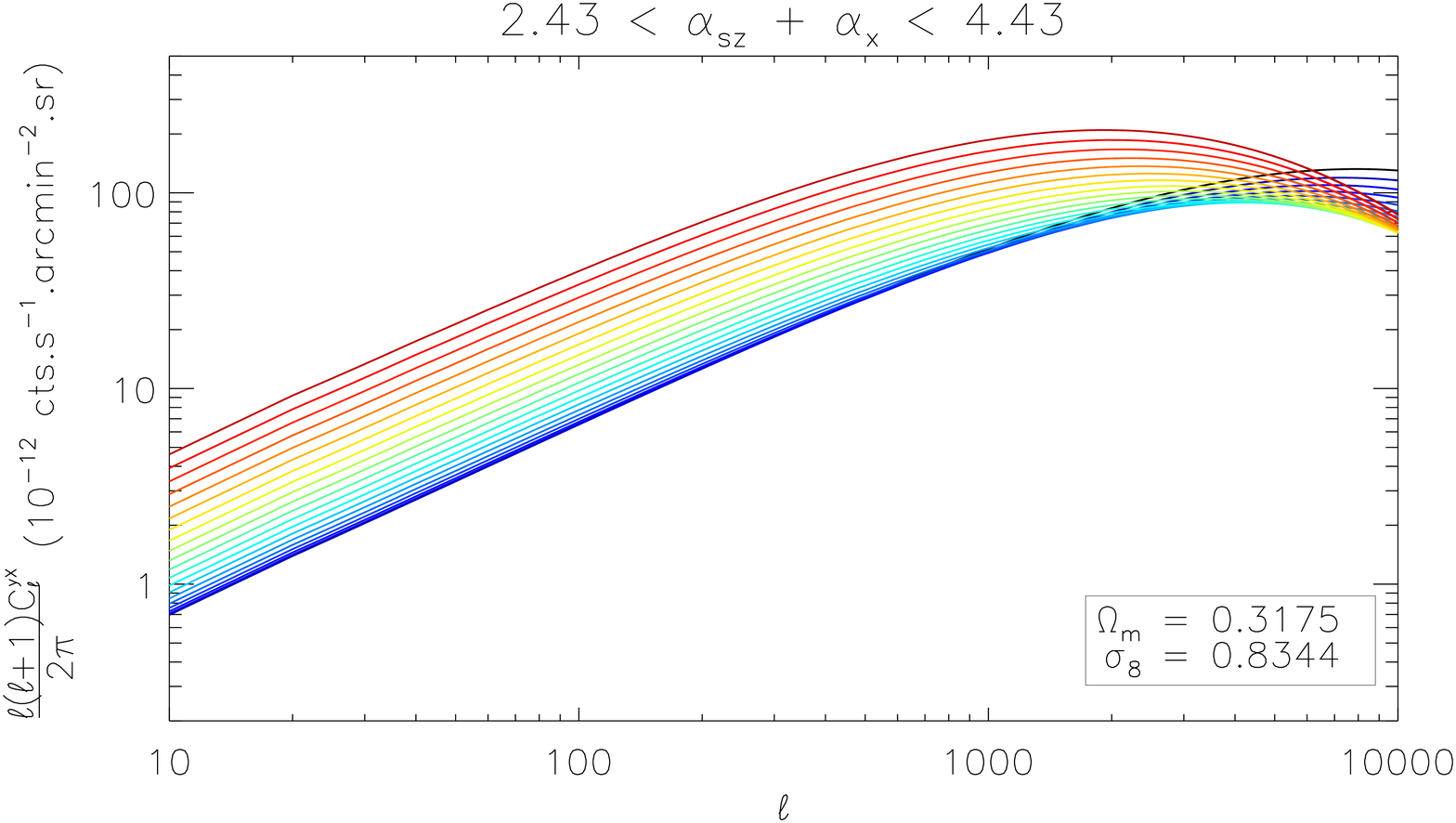}
\includegraphics[scale=0.2]{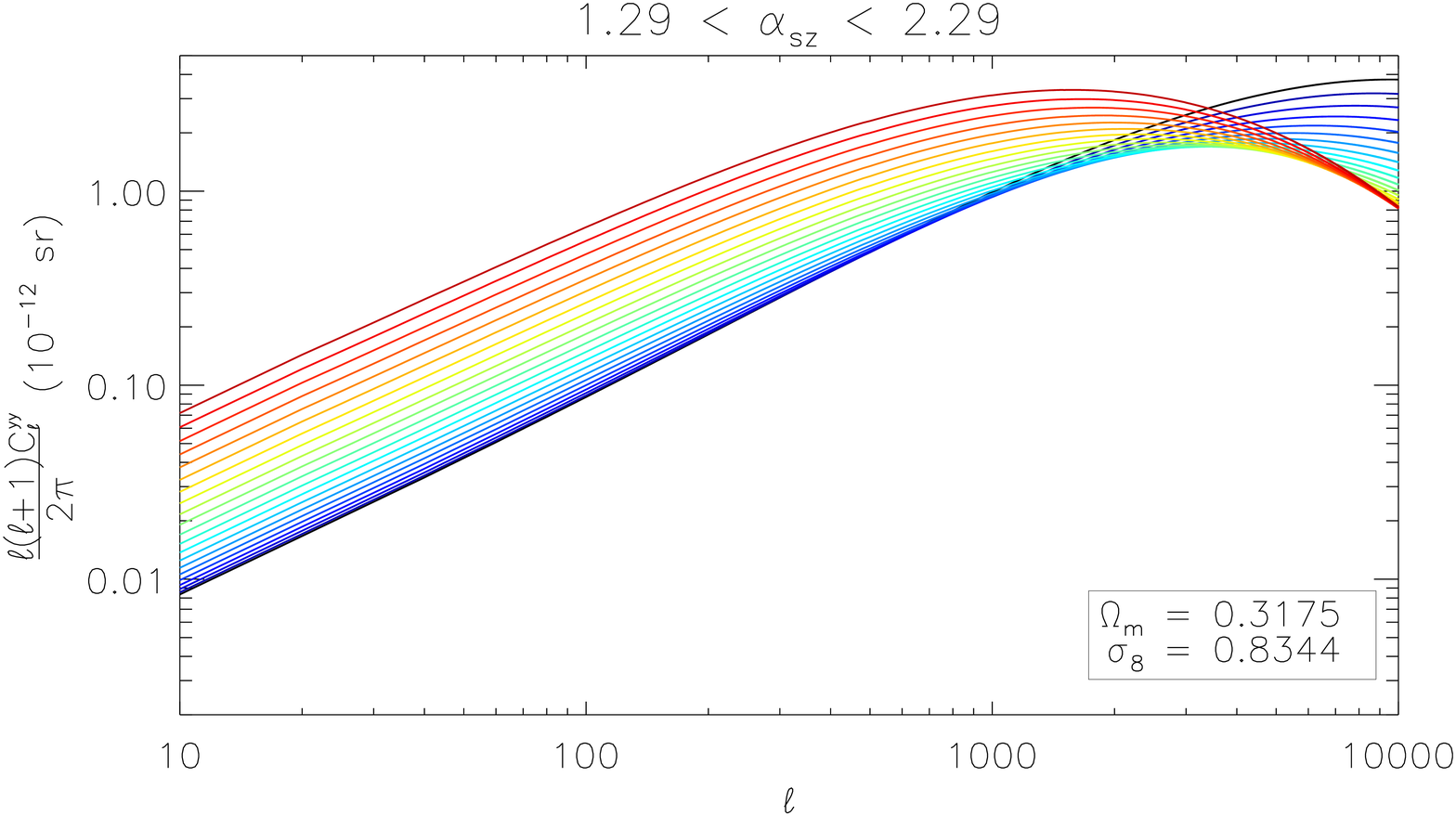}
\includegraphics[scale=0.2]{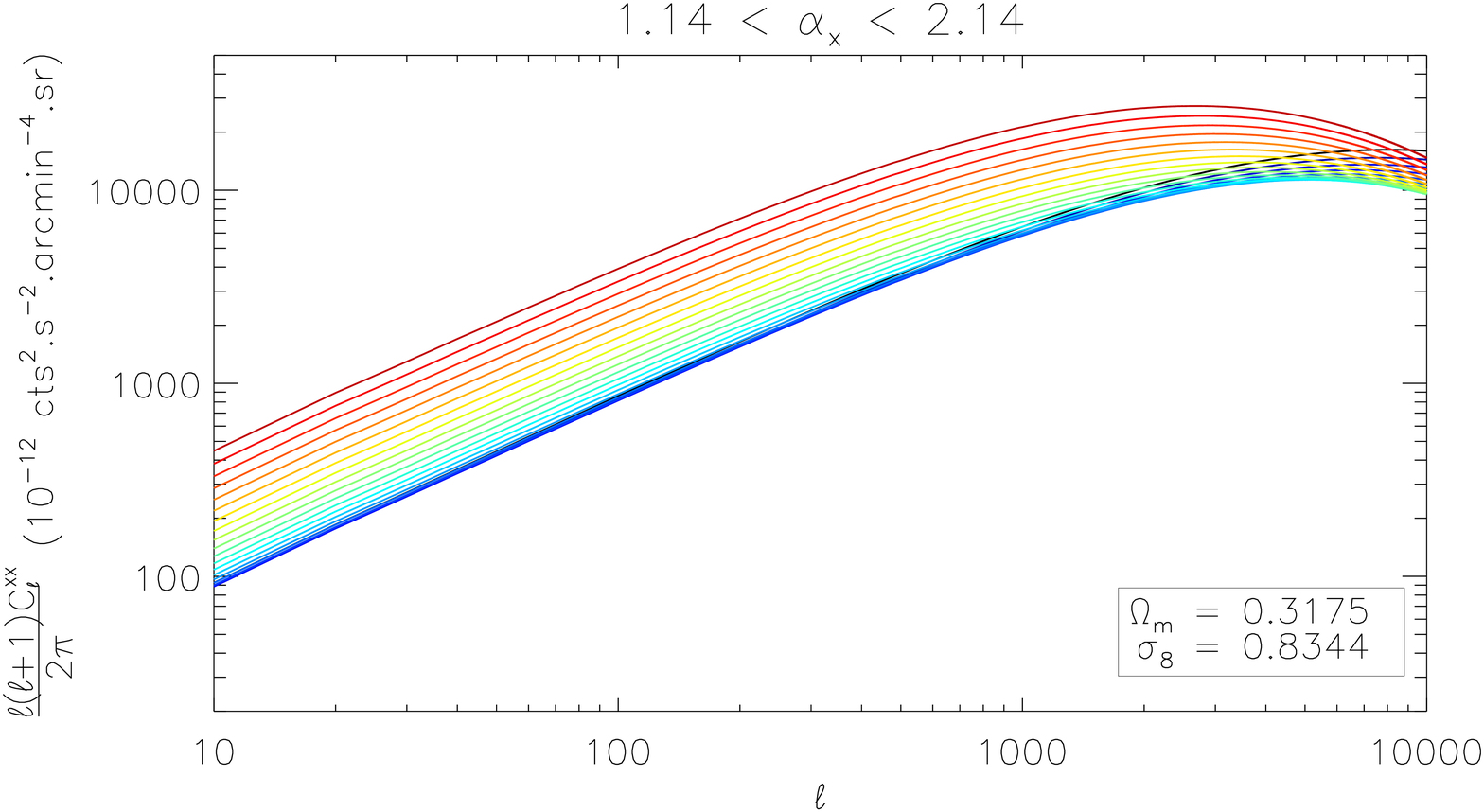}\\
\caption{From top to bottom: Variation of the theoretical tSZ-X, tSZ-tSZ and X-X power spectra as a function of $\alpha_{\rm sz} + \alpha_{\rm x}$, $\alpha_{\rm sz}$ and $\alpha_{\rm x}$, with a step of 0.05, 0.025, and 0.025 respectively.}
\label{slvar}
\end{center}
\end{figure}

In addition to the sensitivity to cosmological parameters, the tSZ-X correlation is highly sensitive to the scaling relations described in Sect.~\ref{secscal}. 
A variation of the scaling relations normalization translates into a variation of the amplitude of the cross spectrum. However, the scaling law power law indexes will produce a modification of the shape of the tSZ-X correlation.\\
In the following, we model the deviation from our reference scaling laws presented in Sect.~\ref{secscal} as
\begin{align}
\tilde{Y}_{500}  = {Y}_{500} \left(\frac{M_{500}}{3.10^{14}} \right)^{\delta \alpha_{\rm sz}} \\
\tilde{L}_{500} = {L}_{500} \left(\frac{M_{500}}{3.10^{14}} \right)^{\delta \alpha_{\rm x}},
\end{align}
where $\delta \alpha_{\rm sz}$ and $\delta \alpha_{\rm x}$ represent the deviations from the reference scaling law indexes, $\alpha_{\rm sz}$ and $\alpha_{\rm x}$, for $M_{500}-Y_{500}$ and $M_{500}-L_{500}$ respectively.\\
In Fig.~\ref{slvar}, we present the variation of the tSZ-X, X-ray and tSZ power spectra with the scaling law indexes, $\alpha_{\rm sz}$ and $\alpha_{\rm x}$, with a step of 0.025 for each index and 0.05 for their sum. We note that the tSZ-X power spectrum is only sensitive to the sum of the indexes, $\alpha_{\rm sz} + \alpha_{\rm x}$.\\
In terms of amplitude, the tSZ-X power spectrum follows
\begin{equation}
A_{\rm cl} \propto  \left(\frac{Y_\star}{0.65}\frac{L_\star}{1.88}\right) \left( \frac{1-b}{0.8}\right)^{\alpha_{\rm sz}+\alpha_{\rm x}}.
\label{modnorm2}
\end{equation}

We observe on Fig.~\ref{slvar}, that increasing the value of $\alpha_{\rm sz}$ and $\alpha_{\rm x}$ increases the power at low-$\ell$ where high-mass objets dominate the signal and decreases the power at high $\ell$ that are dominated by low-mass objects. However, the shape distortion of the power spectra occurs at high-$\ell$. 
The impact of scaling law indexes start to be significant at $\ell > 1000$, $\ell > 800$ and $\ell > 2000$ for tSZ-X, tSZ and X power spectra respectively. We observe that the X-ray power spectrum presents the lowest sensitivity in terms of shape with respect to $\delta \alpha_{\rm x}$.\\

We can infer a global dependence of the tSZ-X cross correlation amplitude,
{\small
\begin{equation}
A_{\rm cl} =  \left(\frac{\sigma_8}{0.83}\right)^{8.12}  \left( \frac{\Omega_{\rm m}}{0.32}\right)^{3.42} \left( \frac{H_0}{67}\right)^{2.38} \left(\frac{Y_\star}{0.65}\frac{L_\star}{1.88}\right) \left( \frac{1-b}{0.8}\right)^{\alpha_{\rm sz}+\alpha_{\rm x}}  \left( \frac{{N}}{{N}_{0}}\right),
\label{modnorm}
\end{equation}
}
where $N$ is a normalization parameter for the mass-function.

\begin{table*}
\center
\caption{Modelling error budget for tSZ-X cross spectrum, X and SZ auto spectra. We notice that the propagation of the uncertainty on $\alpha_{\rm SZ,X}$ depends on the multipole, consequently we provide a range for the uncertainty for $0 < \ell < 2000$. We also propagate the uncertainties of the cosmological parameters to the power spectra, we consider both our fiducial cosmology and the best fitting cosmology from \citet{PlanckSZC}. The total uncertainties is computed assuming {\it cosmo} 2. All uncertainties are expressed in percent of the total power.}
\begin{tabular}{|c|c|c|c|c|c|c|c|c|}
\hline
 Spectrum  & $b$ & $Y_{\star} - {L}_\star$ &$\alpha_{\rm SZ,X}$ & $mf$ & total cosmo fix & {\it cosmo} 1 & {\it cosmo} 2 & Total\\
 \hline
 X-tSZ & 35\% & 9\% & 1-4\% & 10\% & 37\% & 31\% & 27\% & 47\%\\
 tSZ-auto  & 32\% & 9\% & 1-5\% & 10\% & 38\% & 34\% & 28\% & 49\% \\
 X-auto  & 37\% & 16\% & 2-4\% & 10\% & 37\% & 26\% & 26\% & 47\%\\
\hline
\end{tabular}
\label{moderrtab}
\end{table*}

\begin{figure*}[!ht]
\begin{center}
\includegraphics[scale=0.2]{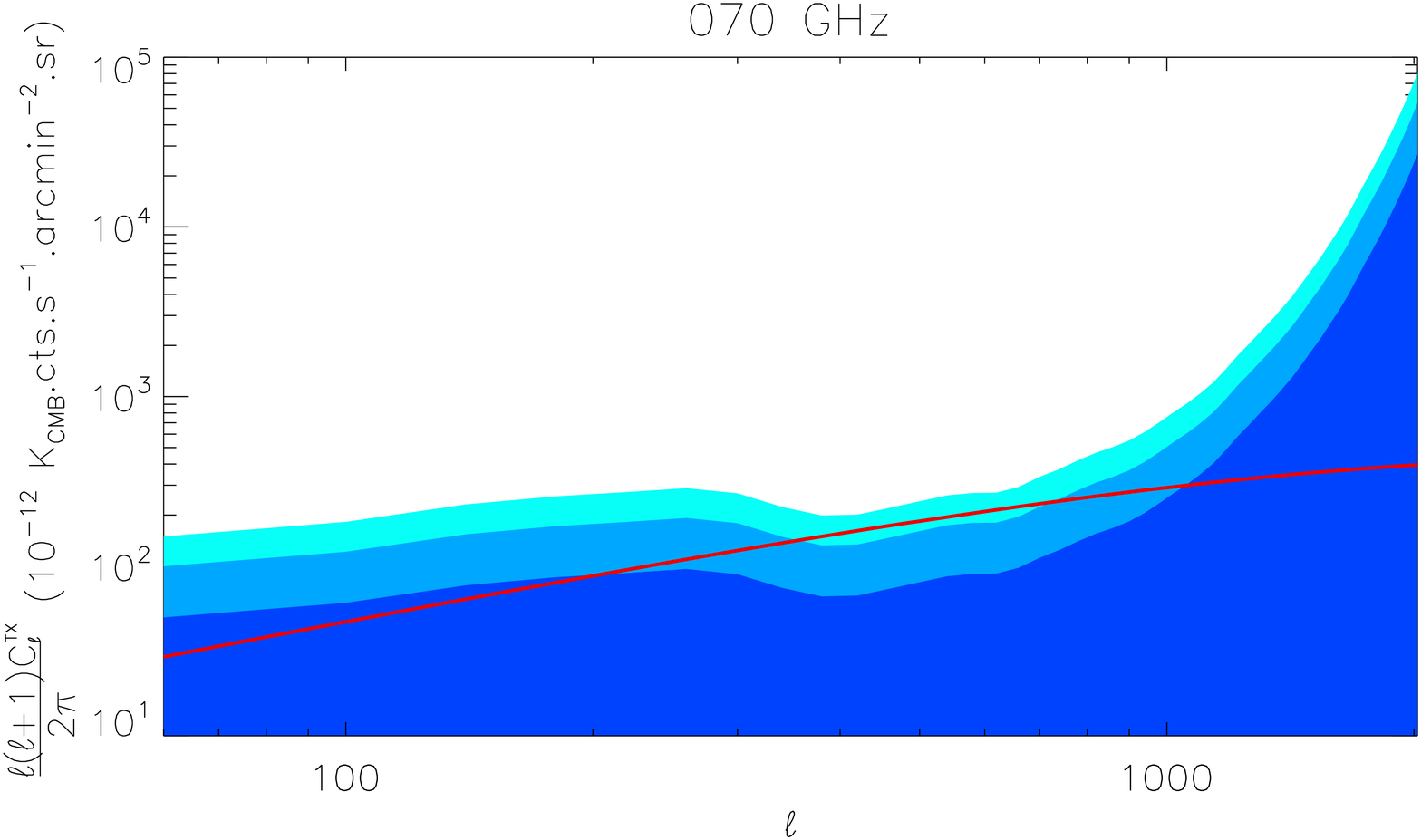}
\includegraphics[scale=0.2]{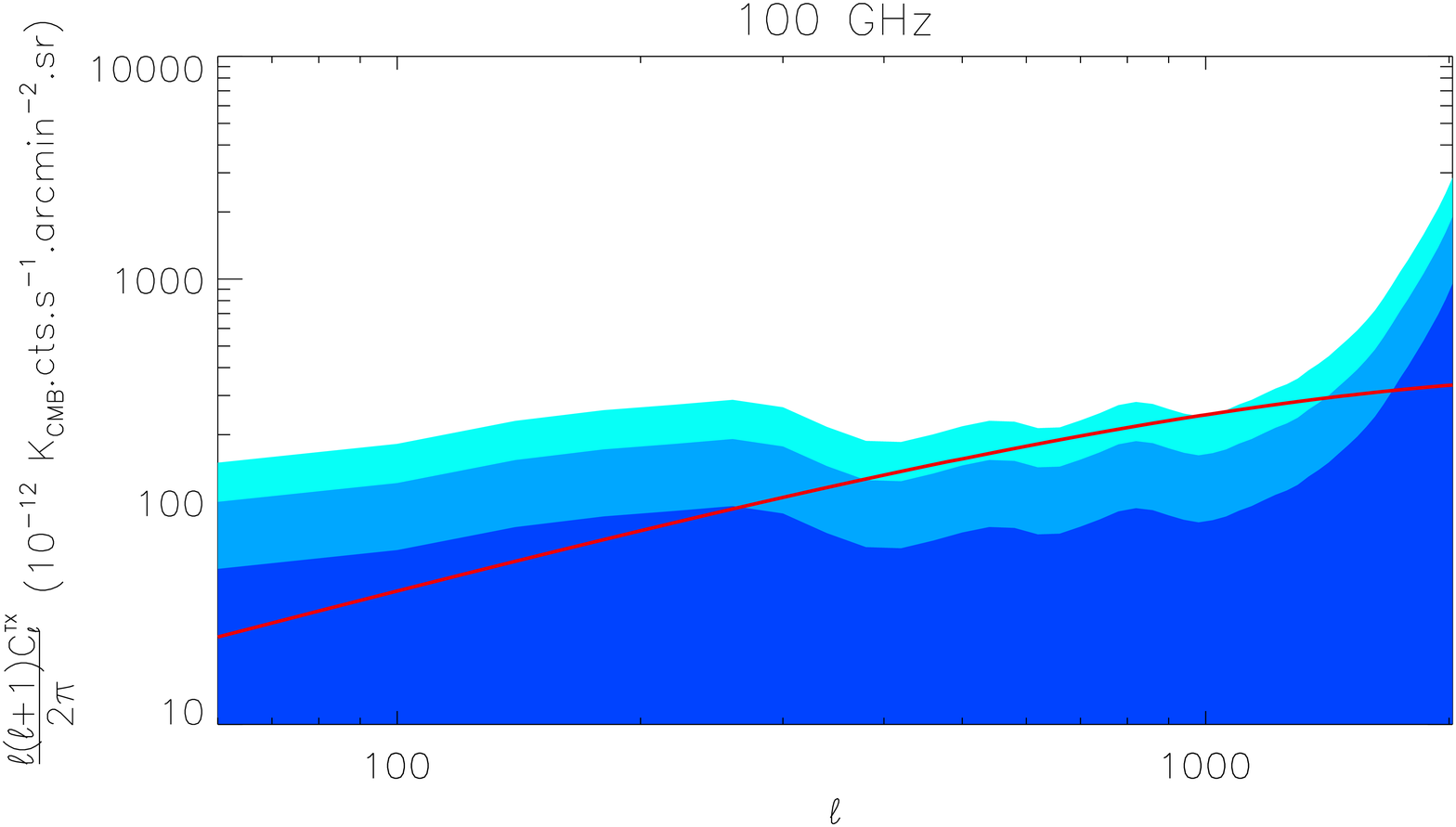}
\includegraphics[scale=0.2]{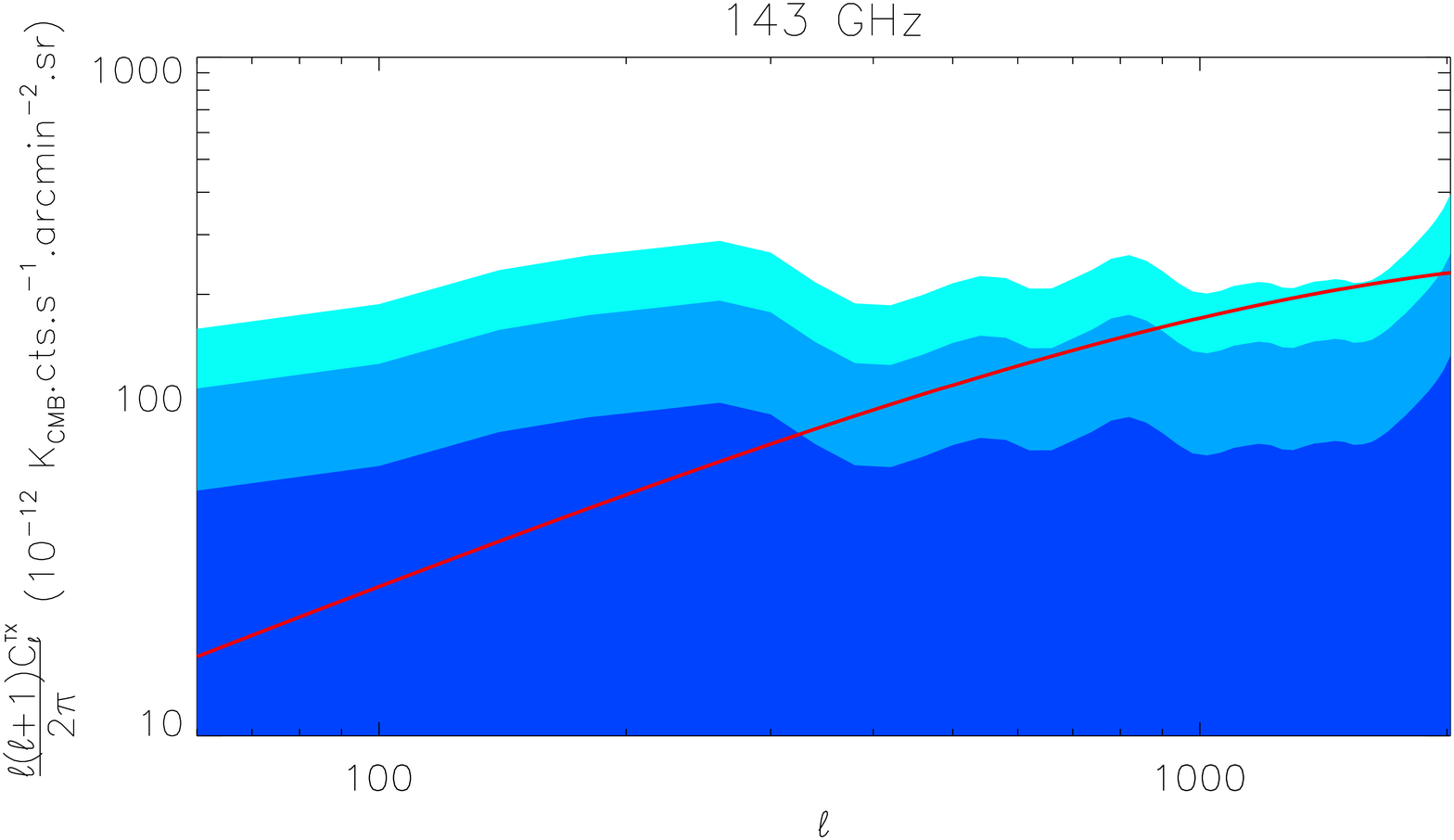}
\includegraphics[scale=0.2]{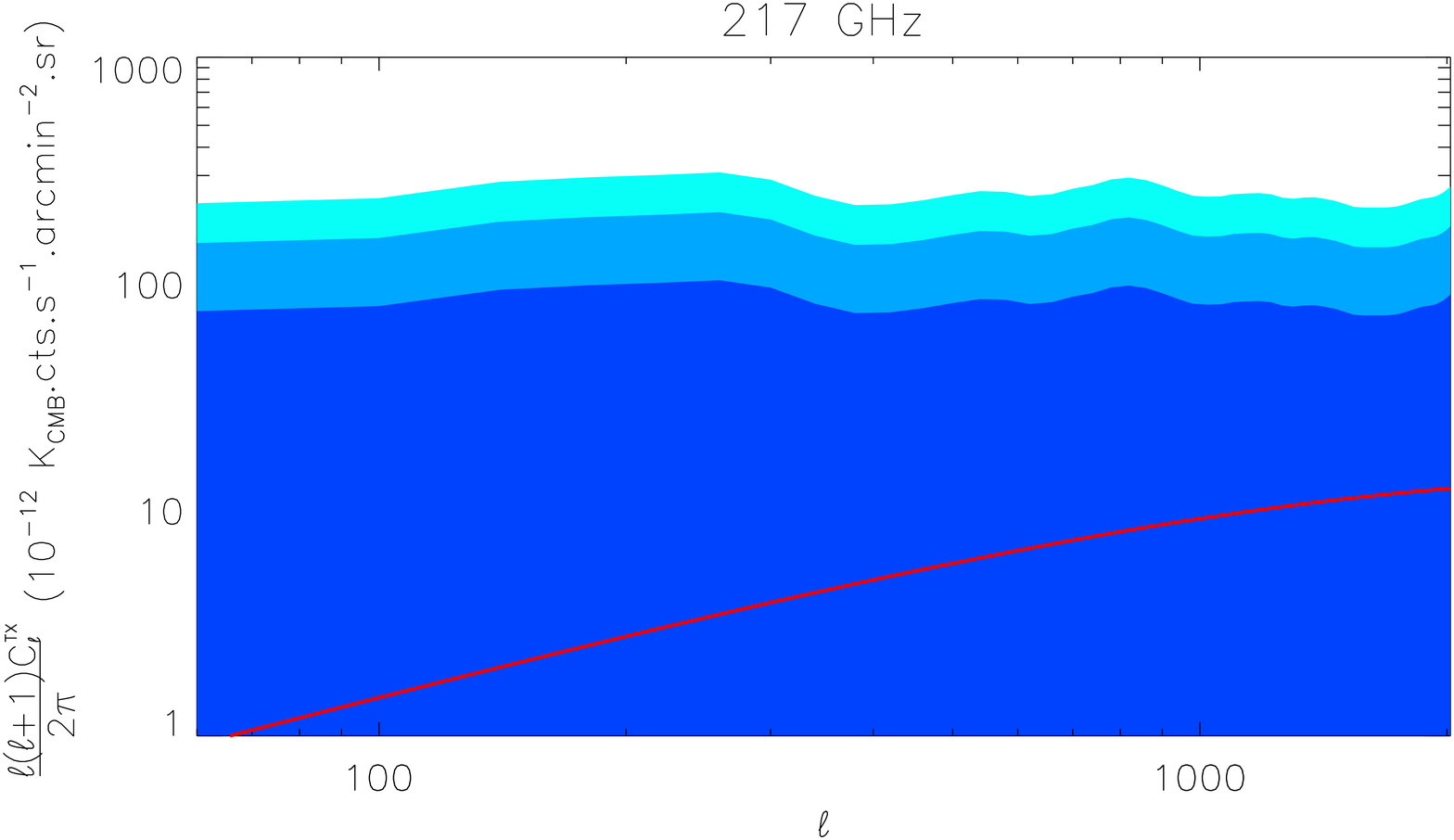}
\includegraphics[scale=0.2]{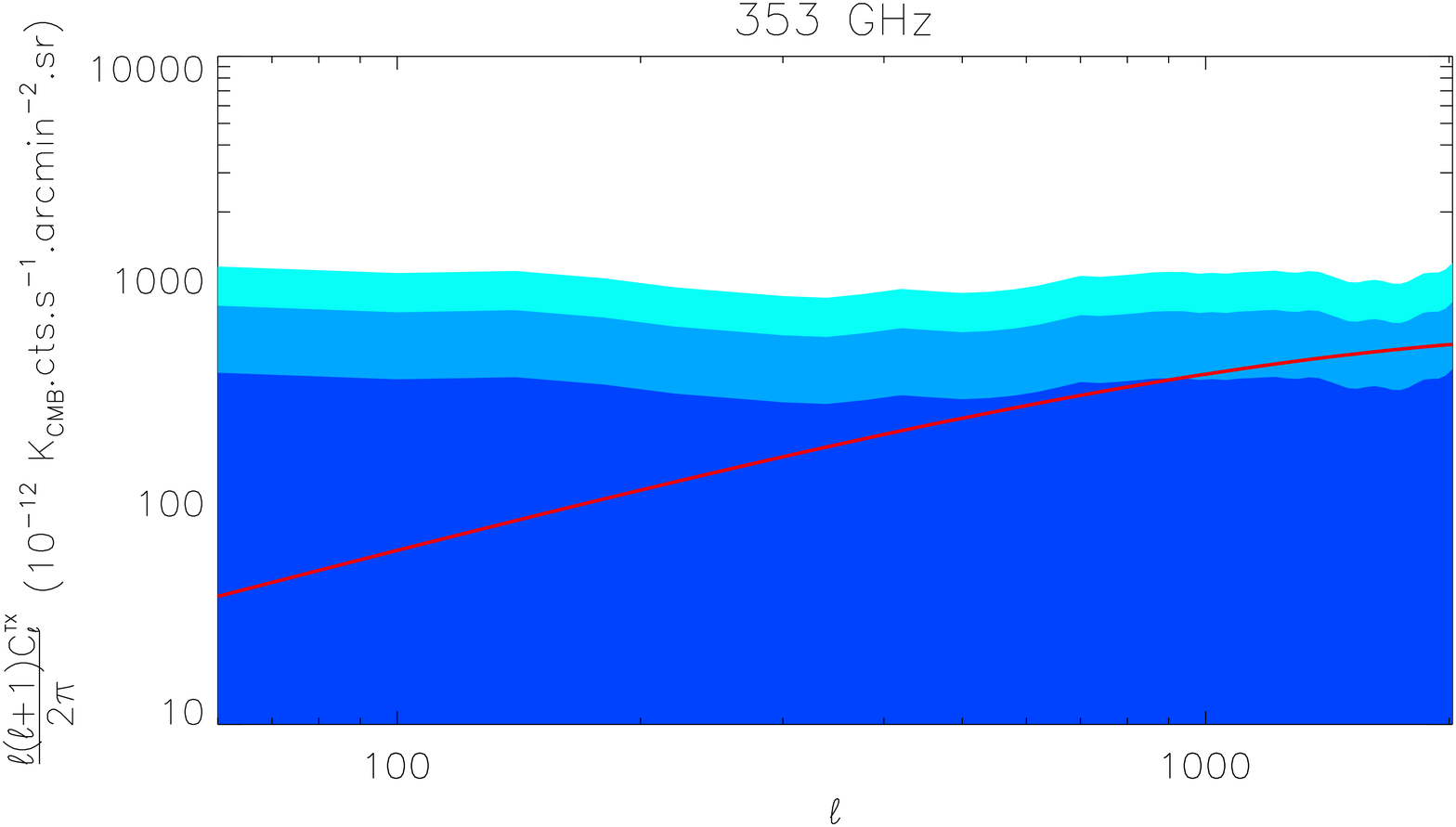}
\includegraphics[scale=0.2]{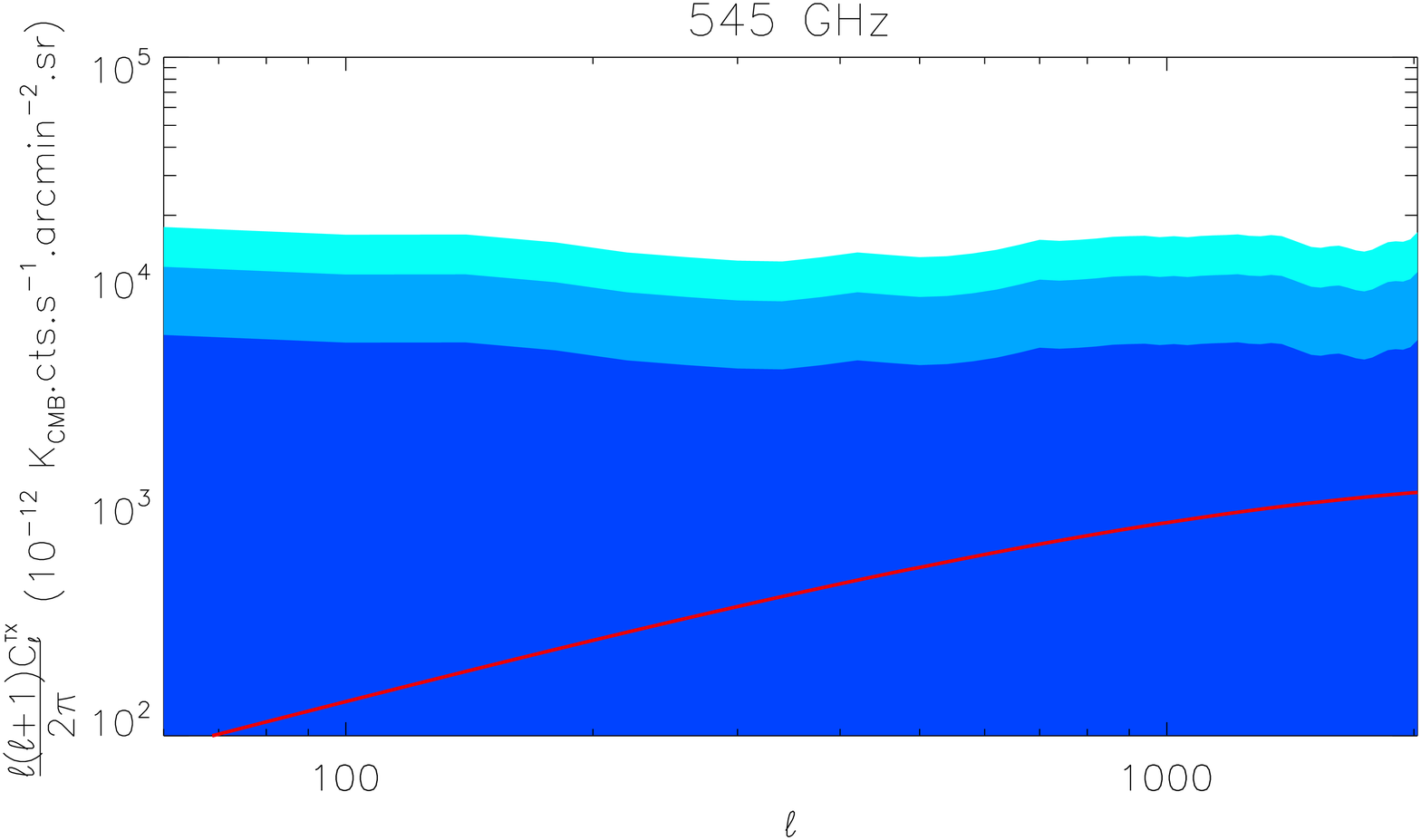}
\caption{From left to right and top to bottom: predicted sensitivity to the tSZ-X cross correlation by computing the cross power spectra between the RASS hard-band and the Planck channels from 70 to 545 GHz. The dark-blue, blue and light-blue shaded regions shows the uncertainties levels at 1, 2, and 3 $\sigma$ for multipole bins $\Delta \ell = 40$. The absolute value of the theoretical tSZ-X cross angular power spectrum is displayed as a red line.}
\label{freqrass}
\end{center}
\end{figure*}

\subsection{Modeling uncertainties}
\label{moderr}
Uncertainties on the predicted spectrum are produced by uncertainties on galaxies clusters properties and uncertainties on the cosmology. 
We used the uncertainties on clusters properties listed in Table~\ref{tabscal} and propagated them to predict the power spectra.
For the mass function, we assumed an over all uncertainty of 10\% \citep{evr02} and for the bias, $b$, we assumed an uncertainty of 10\% \citep[e.g,][]{pif08}.
For the uncertainties on cosmological parameters, we consider two different set of cosmological parameters, our fiducial model, named {\it cosmo} 1, and a second model based on the best fit from \citet{PlanckSZC}, named {\it cosmo} 2. For both set of parameter, we propagated the uncertainties to the tSZ-X power spectra. We carefully account for correlation between uncertainties between parameters for each set.\\

In Table.~\ref{moderrtab}, we present the modeling uncertainties on the tSZ-X cross correlation. We stress that these uncertainties translate into an overall normalization of the spectrum. We also provide the uncertainty levels for the tSZ and X-ray auto correlation spectra.\\ 
Assuming a fixed cosmology, we noted that the error budget is limited by our knowledge on $b$, leading to an uncertainty of about 35\% on the amplitude of the tSX power spectrum.
If we propagate the uncertainties on cosmological parameters to tSZ-X power spectrum, we derived an uncertainty of 31\% for the {\it cosmo} 1 and an uncertainty of 27\% for {\it cosmo} 2. The {{\it cosmo} 2 allows to obtain slightly lower uncertainties for the tSZ-X prediction, because the degeneracy between cosmological parameters is similar for tSZ and tSZ-X spectra, contrary to {\it cosmo} 1 cosmology which presents different degeneracies.\\
We finally note that for the tSZ-X cross power spectrum the error budget from cluster properties and cosmological parameters have the same order of magnitude. The total uncertainty including both contributions reaches 47\%.

\section{Prediction of the tSZ-Xray spectrum measurement}

\label{secpred}
In order to estimate the tSZ-Xray cross correlation, it is possible to use several approaches. One of them relies on the cross correlation of frequency maps of the microwave sky with an X-ray map, see Sect.~\ref{secmeth1} and Sect.~\ref{secmeth2}. Another one consists in using a recovered tSZ map \citep[see e.g.,][]{hur13a} and an X-ray map, see Sect.~\ref{secmeth3}. In the following, we discuss the advantages and drawbacks of each approach.\\
The measurement of the tSZ-X correlation is limited by both, instrumental characteristics and contaminating astrophysical emissions. 
In order to estimate the constraints that can be reached on the tSZ-X power spectrum, we have performed a prediction of the expected signal-to-noise ratio assuming the {\it Planck} nominal mission characteristics \citep[i.e., noise and beams][]{PlanckOVER} for the micro-wave observations and the RASS \citep{RASS} for X-rays.

\subsection{The micro-wave and X-ray skies}

To simulate the sky emission at microwave frequencies, with the appropriate level of noise, we have used the Planck Sky Model \citep[PSM, see][and references therein]{psm}.
At microwave frequencies, the main astrophysical emissions are the diffuse Galactic free-free, synchrotron, thermal dust emissions, the
anomalous microwave emission, the emission from Galactic and extragalactic point sources, the CIB, the Zodiacal light emission, and the tSZ effect in clusters of galaxies.\\

To account for the signal from the X-ray sky we have used the RASS data, in the energy range [0.5,2.0] keV, degraded at an angular resolution of FWHM = 2', reprojected in the HEALpix pixelisation \citep{gor05} following a nearest neighbors interpolation.
There exists several types of astrophysical objects that emit in the X-rays, extra-galactic ones such as galaxy clusters, black holes in AGN, the combination of unresolved X-ray emitting objects produced the X-ray background \citep{fre92}, but also galactic sources mainly supernova remnants and stars.\\

In addition to the signal from galaxy clusters, there is additional astrophysical emissions that are also correlated between the X-ray and the microwave skies. 
This is the case of the radio-loud AGNs and the CIB.\\
Both AGN and CIB present a different frequency dependence than the tSZ effect and consequently can, in principle, be separated from the galaxy-clusters contribution to the tSZ-X cross correlation.
In addition, AGNs are punctual at {\it Planck} and ROSAT angular resolutions, and thus can be separated from the clusters contribution by the shape of the power spectrum.\\
The emission from our galaxy is also present on both sky, due to synchrotron, Free-Free and thermal dust at microwave frequencies and due to $n_{\rm H}$ absorption and galactic X-ray emissions in ROSAT energy bands.
However, considering only ROSAT hard-band (0.5-2.0 keV) reduces the effect of the $n_{\rm H}$ absorption, and utilizing a galactic mask reduces the contamination by galactic foregrounds. 
Similarly to the contamination by extra-galactic point sources, such contamination will present a frequency dependence that differ from a tSZ spectrum, and thus can be discriminated using a multi-frequency analysis.\\

\subsection{Statistical and systematic uncertainties}
\label{secases}
We can estimate the statistical uncertainties, on the tSZ-X correlation, using :
\begin{itemize}
\item our prediction for tSZ-X cross correlation power spectrum from galaxy cluster, $C^{yx}_\ell$,
\item our prediction for tSZ and X-ray auto-correlation power spectra from galaxy cluster, $C^{yy}_\ell$ and $C^{xx}_\ell$,
\item the measured cross correlation, $C^{T_{\nu} {\rm R}}_\ell $, between the microwave sky from the PSM, noted $T_{\nu}$ for frequency $\nu$, and the X-ray sky from RASS data, noted ${\rm R}$,
\item the measured auto correlations, $C^{T_{\nu} T_{\nu}}_\ell $ and $C^{{\rm R} {\rm R}}_\ell $, of the microwave and X-ray skies,
\item the measured cross correlation, $C^{y_{\rm PSM} {\rm R}}_\ell $, between the tSZ map constructed from the PSM, noted $y_{\rm PSM}$, and the X-ray sky from RASS data, noted ${\rm R}$,
\item the measured auto correlation, $C^{y_{\rm PSM} y_{\rm PSM}}_\ell $ of the tSZ map.
\end{itemize}
The expression uncertainties for each tSZ-X detection methods is presented in Sect.~\ref{secmeth1}, Sect.~\ref{secmeth2} and Sect.~\ref{secmeth3}.\\

Our estimation of uncertainties through simulations of the microwave sky does not allows to estimate the systematic uncertainties. 
In order to account for systematics uncertainties we consider three cases for the description of the measured tSZ-X cross correlation.
\begin{itemize}
\item {\it Case} 1 : considering only the contribution of galaxy clusters to the tSZ-X correlation.
\item {\it Case} 2 : considering the contributions from galaxy clusters and AGNs to tSZ-X correlation.
\item {\it Case} 3 : considering galaxy clusters, AGNs, and CIB-X contributions.
\end{itemize}
The complete description of the cross power spectrum for the case 3 reads,
\begin{equation}
C_\ell = \left[A_{\rm cl}\, g(\nu) + A_{\rm CIB}\, F_{\rm CIB}(\nu)\right] C^{yx}_\ell + A_{\rm AGN} \, F_{\rm AGN}(\nu),
\end{equation}
with $F_{\rm CIB}$ the CIB SED \citep{gis00}, $F_{\rm AGN}(\nu)$ a typical radio-loud AGN SED assuming a spectral index of -0.7 in intensity units, $A_{\rm cl}$ (see Eq.~\ref{modnorm}), $A_{\rm CIB}$, and $A_{\rm AGN}$ are the parameters of the model and account respectively for galaxy clusters, CIB-X, and AGN contributions to the tSZ-X cross power spectrum. The {\it case} 1 assumes $A_{\rm CIB} = 0$ and $A_{\rm AGN} = 0$ and the {\it case} 2 assumes $A_{\rm CIB} = 0$.\\
This modeling assumes that the CIB-X correlation presents a similar shape, as a function of $\ell$, than the clusters contribution to tSZ-X cross spectrum and that the AGN contribution is Poissonian, as the AGN clustering can be neglected.\\
In the following, we consider the multipole range $40 < \ell < 2000$ for our signal-to-noise ratio prediction.

\subsection{Cross correlation spectrum from frequency maps}
\label{secmeth1}

The tSZ-X cross correlation can be directly estimated from the correlation between X-ray maps and microwave full-sky observations at a given frequency, noted $C^{\nu R}_\ell$. 
We estimate the expected level of uncertainties when correlating the RASS hard-band and microwave maps at frequency $\nu$.\\
We mask 30\% of the sky by applying a cut on the thermal dust emission intensity. Then, we estimate the uncertainties following
\begin{align}
\label{eqcl}
 \left(\Delta C^{\nu R}_\ell \right)^2 =& \frac{g(\nu)^2 \left[ \left(C^{yx}_\ell \right)^2 +  C^{yy}_\ell C^{xx}_\ell \right]}{(2\ell +1)f_{\rm sky}} \nonumber \\
	&+  \frac{\left(C^{T_{\nu} {\rm R}}_\ell \right)^2}{(2\ell +1)f_{\rm sky}}  \\
	&+  \frac{\left(C^{T_{\nu} T_{\nu}}_\ell + g(\nu)^2 C^{yy}_\ell \right) \left(C^{\rm RR}_\ell - C^{xx}_\ell  \right)}{(2\ell +1)f_{\rm sky}}, \nonumber
\end{align} 
where $\Delta C^{\nu R}_\ell$ is the uncertainties on the cross power spectrum between microwave and X-ray skies, $f_{\rm sky}$ is the sky fraction used for the analysis.\\
The first term in Eq.~\ref{eqcl} corresponds to the cosmic variance of the tSZ-X cross correlation, the other terms account for the uncertainties produced by foreground emissions.\\

In Fig.~\ref{freqrass}, we present the resulting uncertainty level at 1, 2 and 3 $\sigma$ as a function of the frequency. We also present the expected absolute value of the tSZ-X cross correlation for our fiducial model. 
All spectra are displayed in units of ${\rm K}_{\rm CMB}.{\rm cts.s}^{-1}.{\rm arcmin}^{-2}.{\rm sr}$. Each spectrum has been corrected for the mask effect \citep[see][]{tri05} and the beam effects. 
We choose to present the uncertainty for multipole bins with $\Delta \ell = 40$.\\
For the lowest frequencies, below 70~GHz, the signal is completely dominated by the instrumental noise contribution. 
For intermediate frequencies, from 70 to 217~GHz, the main uncertainty is the CMB contamination, the uncertainty level clearly shows the CMB features, mainly the first three acoustic pics. 
Above 353~GHz, the uncertainties are dominated by the thermal dust contamination.\\
We note that at 217~GHz the tSZ emission is not rigorously null, however the tSZ transmission in this channel is faint \citep{hur13b}. Consequently, this channel can be used to check systematic effects.\\

In the Table.~\ref{fitfreqr}, we present the expected signal-to-noise for tSZ-X correlation as a function of frequency. The signal-to-noise is provided assuming the {\it cases} 1 and 2.\\
We observe that in a {\it case 3}, we reach a signal-to-noise above 6 for only two channels, 100 and 143~GHz. However, the main limitations in that case is the astrophysical emission from the microwave sky such as CMB and thermal dust emission. These emissions are correlated from a frequency to an other. 

\begin{table*}
\center
\caption{Detection signal-to-noise of the tSZ-X cross power spectrum frequency per frequency from 30 to 857 GHz considering values of $\ell$ from 40 to 2000.}
\begin{tabular}{|c|c|c|c|c|c|c|c|c|}
\hline
Frequency (GHz) & 30 & 44 & 70 & 100 & 143 & 217 & 353 & 545 \\
\hline
{\it Case} 1 & 3.4 & 4.5 & 8.9 & 13.9 & 15.2 & 0.8 & 7.7 & 1.2 \\
{\it Case} 2 & 1.4 & 1.8 & 4.2 & 6.4 & 6.3 & 0.3 & 3.2 & 0.5 \\
\hline
\end{tabular}
\label{fitfreqr}
\end{table*}

\subsection{Cross power spectrum from cleaned frequency maps}
\label{secmeth2}
\begin{figure*}[!th]
\begin{center}
\includegraphics[scale=0.2]{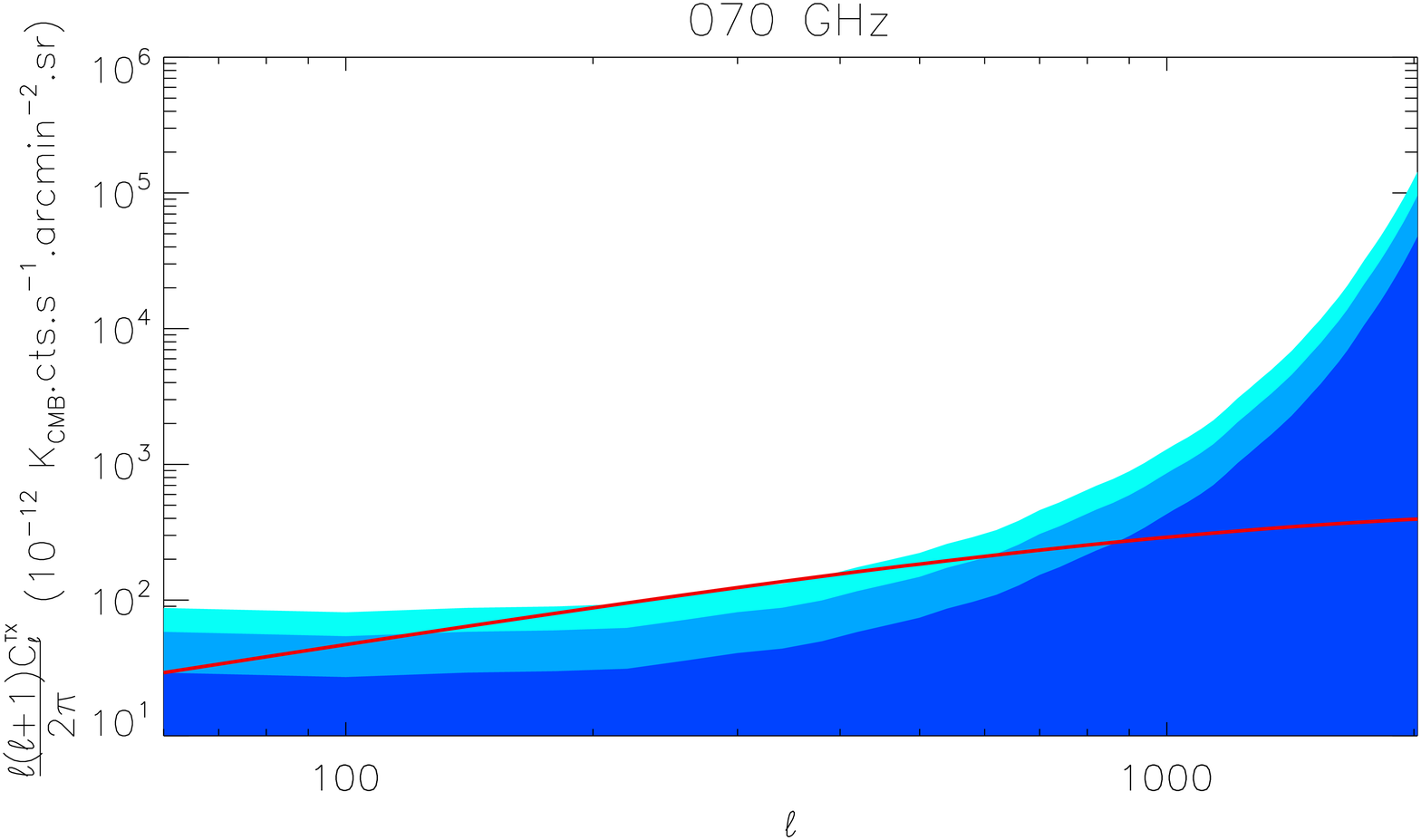}
\includegraphics[scale=0.2]{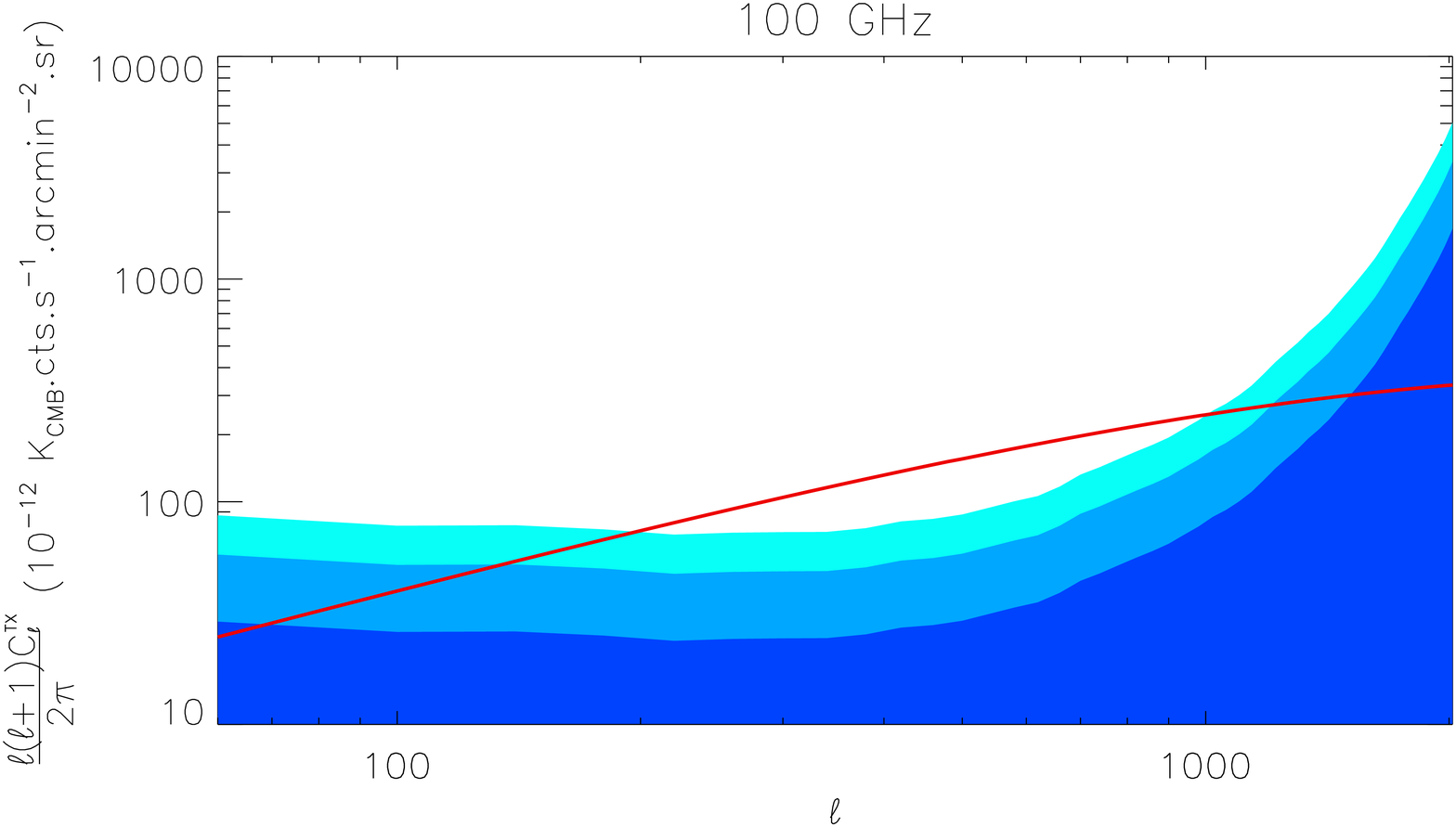}
\includegraphics[scale=0.2]{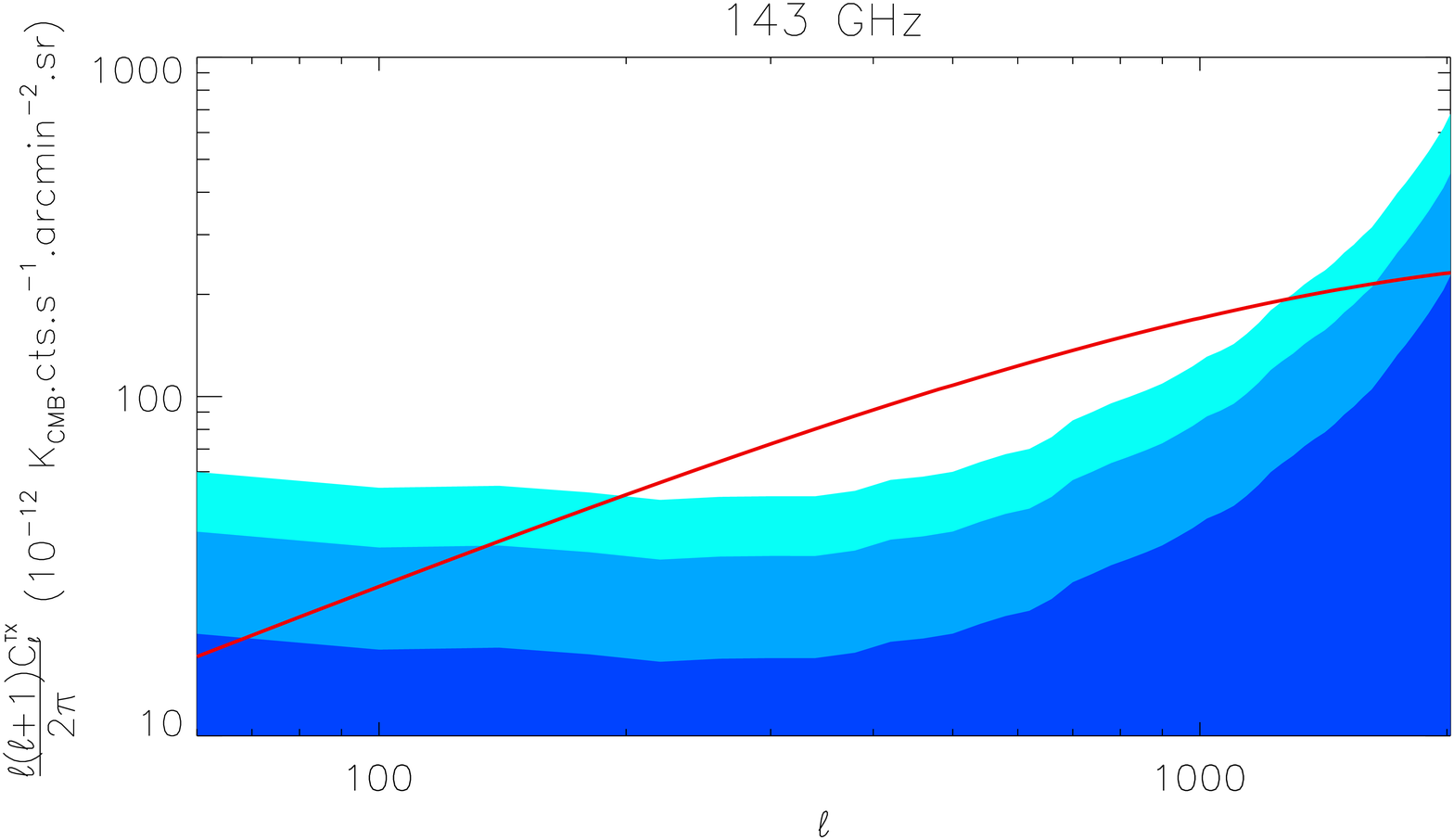}
\includegraphics[scale=0.2]{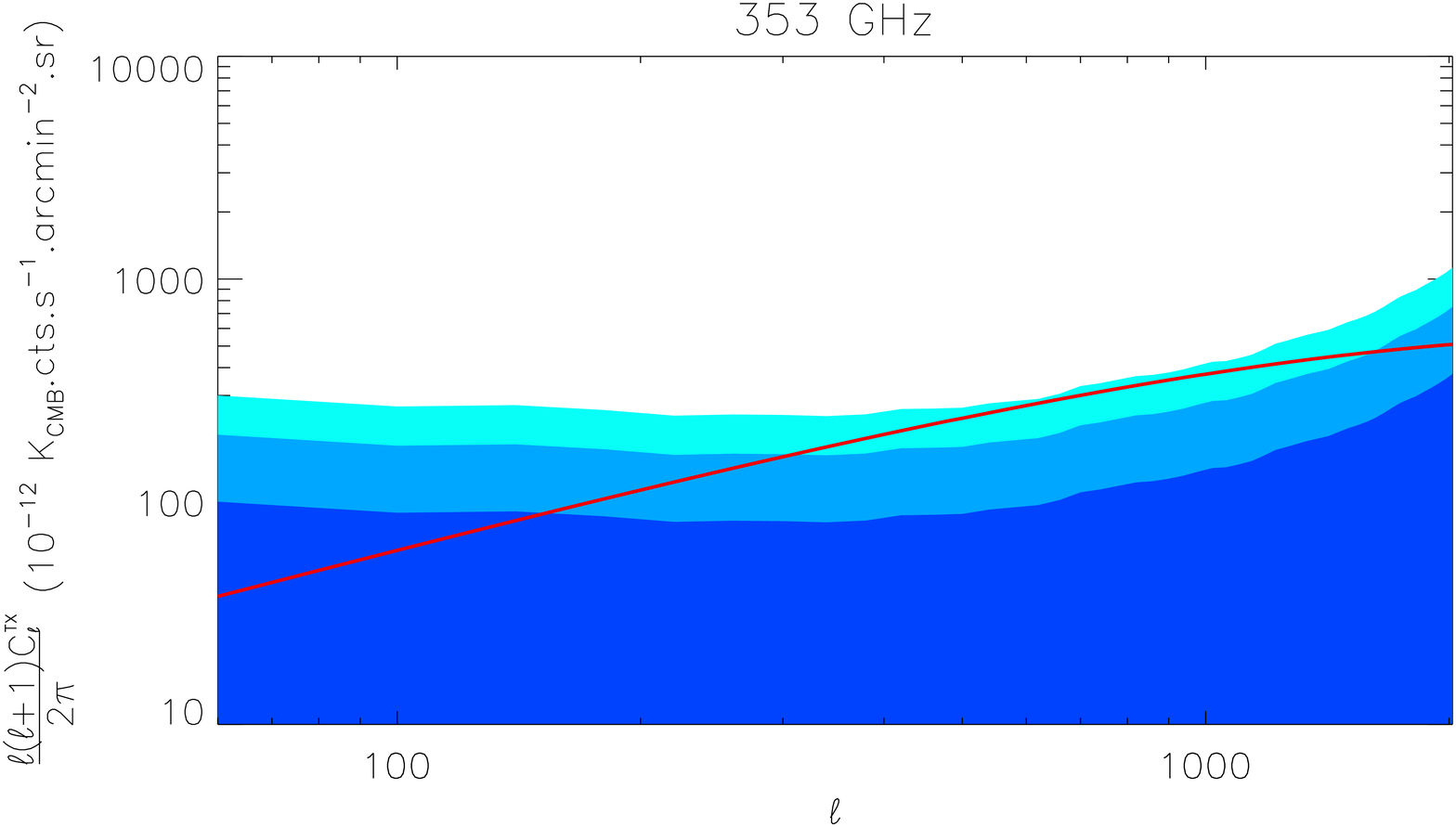}
\includegraphics[scale=0.2]{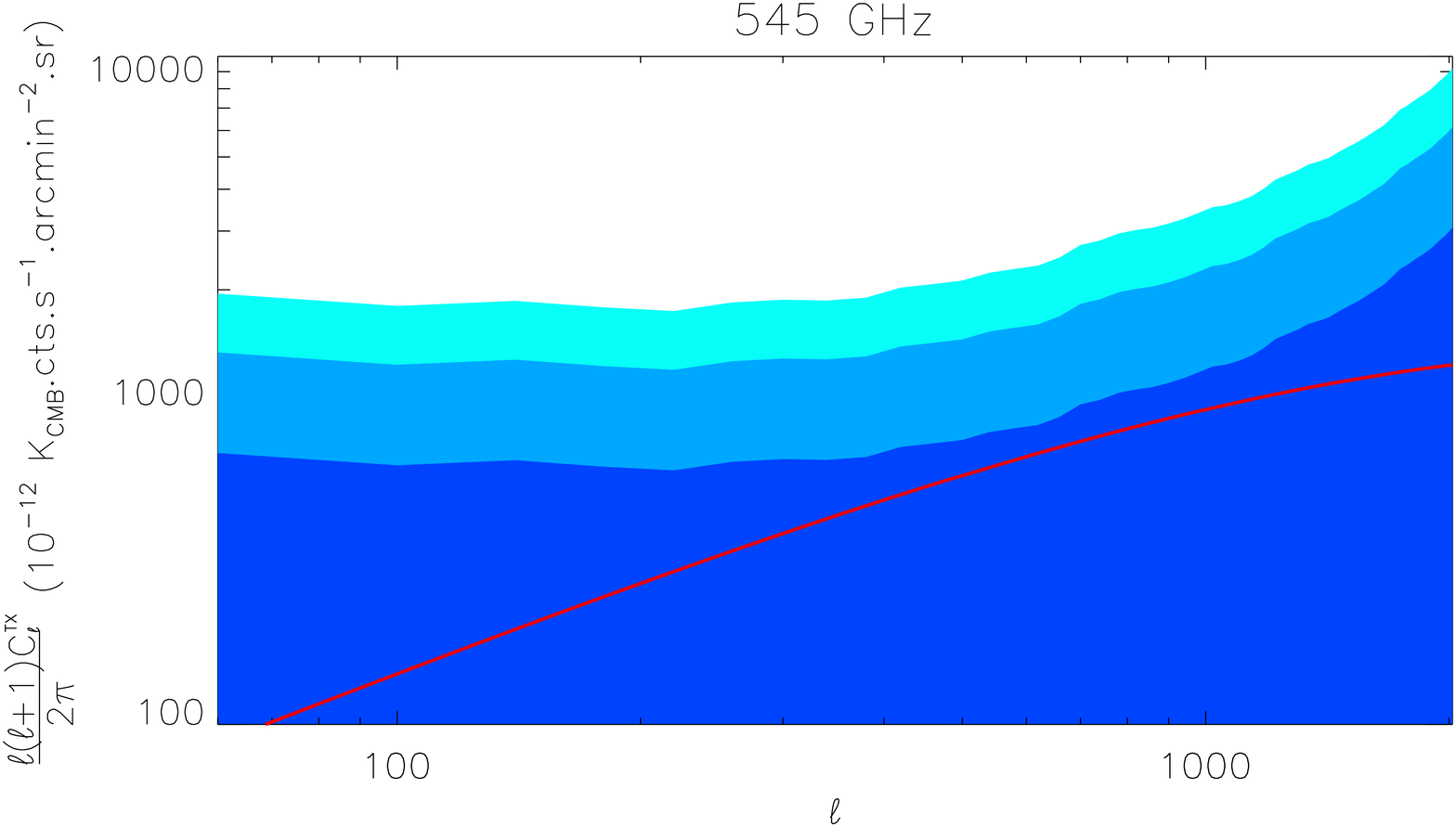}
\caption{From left to right and top to bottom: predicted sensitivity to the tSZ-X cross correlation for the cleaned cross power spectra between the RASS hard-band and the Planck channels at 70, 100, 143, 353, and 545 GHz respectively. The dark-blue, blue and light-blue shaded regions shows the uncertainties levels at 1, 2, and 3 $\sigma$ for multipole bins $\Delta \ell = 40$. The absolute value of th theoretical tSZ-X cross angular power spectrum is displayed as a red line.}
\label{cleanrass}
\end{center}
\end{figure*}
\begin{table*}
\center
\caption{Detection signal-to-noise of the tSZ-X cross power spectrum frequency per frequency after cleaning at 70, 100, 143, 353 and 545 GHz considering values of $\ell$ from 40 to 2000.}
\begin{tabular}{|c|c|c|c|c|c|c|}
\hline
Frequency (GHz) & 70 & 100 & 143 & 353 & 545 & all\\
\hline
{\it Case} 1 & 8.5 & 16.3 & 19.8 & 13.5 & 3.9 & --- \\
{\it Case} 2 & 3.9 & 7.1 & 8.9 & 6.0 & 1.8 & --- \\
{\it Case} 3 & 4.7 & 9.4 & 13.4 & 11.8 & 3.0 & 17.5 \\
\hline
\end{tabular}
\label{fitcleanerr}
\end{table*}
In order to increase the signal to noise of the tSZ-X detection, we combine the different frequencies to remove the contribution from CMB and thermal dust. This cleaning is performed by subtracting the 217~GHz spectrum to other spectra to remove CMB contamination, and decorrelating each channel from the 857 GHz map to reduce thermal dust contamination.\\
\begin{equation}
\tilde{C}_\ell^{T_{\nu} {\rm R}} = {C}_\ell^{T_{\nu} {\rm R}}  - {C}_\ell^{T_{217} \,{\rm R}}  - (\rho_{\nu} - \rho_{217}) {C}_\ell^{T_{857} \, {\rm R}} 
\end{equation}
Where $\rho_{\nu}$ is the correlation factor between the map at the frequency $\nu$ and the map at 857 GHz. This factor is computed on the area of the sky that is not masked.
We compute these cleaned angular power spectra at 70, 100, 143, 353 and 545~GHz.\\
We propagate the uncertainties considering the correlation between cross spectra,
{\small
\begin{align}
< C^{\nu {\rm R}}_\ell,C^{\nu' {\rm R}}_\ell > =& \frac{g(\nu)g(\nu') \left[ \left(C^{yx}_\ell \right)^2  + C^{yy}_\ell C^{xx}_\ell \right]}{(2\ell +1)f_{\rm sky}} \nonumber \\
	&+  \frac{ C^{T_{\nu} {\rm R}}_\ell C^{T_{\nu'} {\rm R}}_\ell }{(2\ell +1)f_{\rm sky}}  \\
	&+  \frac{ \left( C^{T_{\nu} T_{\nu'}}_\ell + g(\nu)g(\nu') C^{yy}_\ell \right) \left(C^{\rm RR}_\ell - C^{xx}_\ell  \right)}{(2\ell +1)f_{\rm sky}}, \nonumber
\end{align}
}
In Fig.~\ref{cleanrass}, we present the obtained power spectrum. We observe that the uncertainties at low-$\ell$ are dominated by foreground residuals. 
Indeed, the decorrelation from the 857~GHz assuming a single scaling coefficient $\rho_\nu$ does not account for thermal dust SED variation across the sky and thus leads to residual emission that dominates the uncertainties at those scales. At high-$\ell$ the uncertainties are dominated by the instrumental noise. In Fig.~\ref{cleanrass}, we also observe that our cleaning is particularly efficient at intermediate scales from $\ell=100$ to $\ell = 1000$, due to CMB contamination removal.\\

In the Table.~\ref{fitcleanerr} we present the expected signal-to-noise of the tSZ-X cross power spectrum signal in the multipole range $40 < \ell < 2000$. We provide the results in the three cases described in Sect.~\ref{secases}.\\
For the {\it case} 2, we performed the adjustment of $A_{\rm AGN}$ individually per frequency.\\
For the {\it case} 3, we performed the estimation of the expected signal-to-noise ratio per frequency and considering all frequencies. We note that for both adjustments the parameters $A_{\rm CIB}$ and $A_{\rm AGN}$ are fitted considering all frequencies. This adjustment is performed considering the global covariance matrix of all spectra.\\
This procedure explains the increase of signal-to-noise between {\it case} 2 and {\it case} 3. As for the {\it case} 3, we consider the multi-frequency information of the tSZ-X cross correlation for the fit of our foreground model.\\
In a realistic case, {\it case} 3, we reach a signal-to-noise of 13.4 for the tSZ-X cross correlation at 143~GHz. Considering all five frequencies, we obtain a global signal-to-noise of 17.5.  

\subsection{Cross correlation from Compton parameter maps}
\label{secmeth3}
\begin{figure}[!th]
\begin{center}
\includegraphics[scale=0.2]{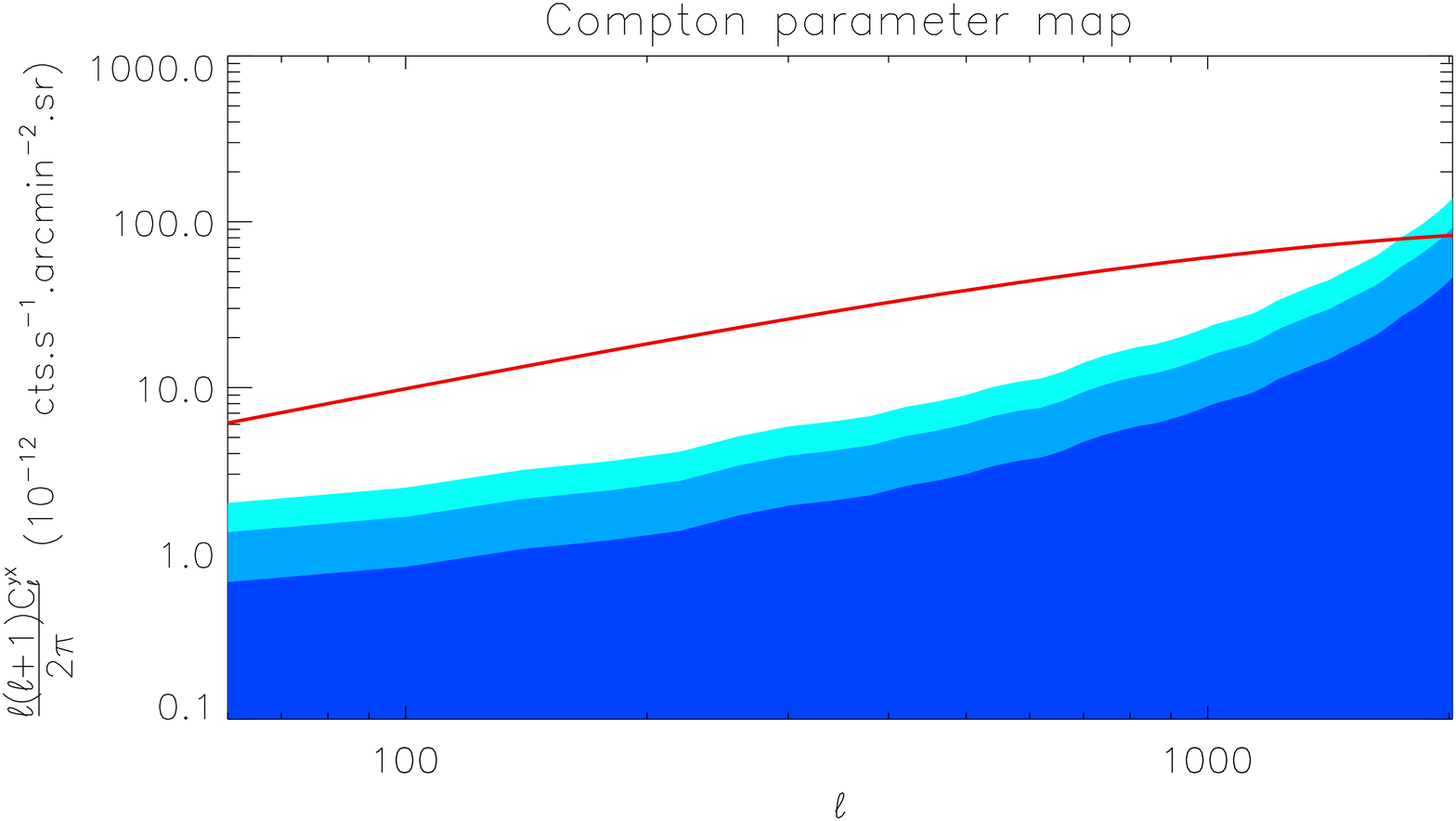}
\caption{Predicted sensitivity to the tSZ-X cross correlation for the cross power spectra between the RASS hard-band and a MILCA tSZ-map. The dark-blue, blue and light-blue shaded regions shows the uncertainties levels at 1, 2, and 3 $\sigma$ for multipole bins $\Delta \ell = 40$. The absolute value of th theoretical tSZ-X cross angular power spectrum is displayed as a red line.}
\label{yrass}
\end{center}
\end{figure}
We construct a tSZ map from the PSM simulations of microwave sky observations from 100 to 857 GHz using the MILCA method \citep{hur13a}. 
Then, we estimate the uncertainties on the cross power spectrum, $C^{y{\rm R}}_\ell$, between a tSZ map and an X-ray map as
\begin{align}
 \left(\Delta C^{y{\rm R}}_\ell \right)^2 =& \frac{\left(C^{yx}_\ell \right)^2 +  C^{yy}_\ell C^{xx}_\ell}{(2\ell +1)f_{sky}} \nonumber \\
	&+  \frac{\left(C^{ y_{\rm PSM}{\rm R}}_\ell \right)^2+\left( C^{y_{\rm PSM}\,y_{\rm PSM}}_\ell +  C^{yy}_\ell \right) \left(C^{\rm RR}_\ell - C^{xx}_\ell  \right)}{(2\ell +1)f_{sky}}, 
\end{align} 
where $y_{\rm PSM}$ corresponds to the tSZ map constructed from the PSM simulations.\\
In Fig.~\ref{yrass}, we present the obtained levels of uncertainties. The tSZ-X power spectrum for our fiducial model is above 2 $\sigma$ for each bin of $\Delta \ell = 40$ from $\ell = 40$ to $2000$. The main limitations are the instrumental noise and CIB residuals that cannot be removed by a linear combination.\\
If we do not consider contamination by correlated astrophysical emissions such as radio-loud AGNs and CIB, we obtain an overall signal-to-noise of 62.3 from $\ell = 40$ to $\ell = 2000$. If we consider contamination by AGNs and CIB we obtain a signal-to-noise of 31.5.\\
We note, in Sect.~\ref{moderr}, that the modeling derived from present constraints leads to about 47\% uncertainty on the amplitude of the tSZ-X cross correlation. 
Considering {\it case} 3 and a signal-to-noise of 31.5,  the amplitude of the tSZ-X cross-correlation can be obtained at 3.2\% precision.
As a consequence, the utilization of a tSZ map allows to set the tighter constraints on the tSZ-X cross-correlation, and allows to increase our knowledge of cosmological and astrophysical parameters in the related degeneracy space by a factor of 15.

\subsection{Constraints on astrophysical and cosmological parameters}
\label{constr}

The amplitude of the tSZ-X cross correlation can be constraints at a precision of 3.2\%. However, this amplitude is sensitive to several parameters, from both cosmology and scaling laws, see Eg.~\ref{modnorm}.
The expected constraints on the tSZ-X cross correlation normalization reads
\begin{equation}
\Delta A_{\rm cl} =  0.03,
\end{equation} 
in the case of the correlation between a tSZ map and an X-ray map considering the contamination by both AGN and CIB emissions.\\
To obtain the uncertainty over a single parameter, it is needed to propagate carefully the uncertainties from other parameters considering the global covariance matrix of all parameters.

Beyond the amplitude of the tSZ-X cross-correlation, the expected high level of significance of the tSZ-X cross correlation should allow us to constrain the scaling-law spectral indexes.\\
\begin{figure}[!th]
\begin{center}
\includegraphics[scale=0.2]{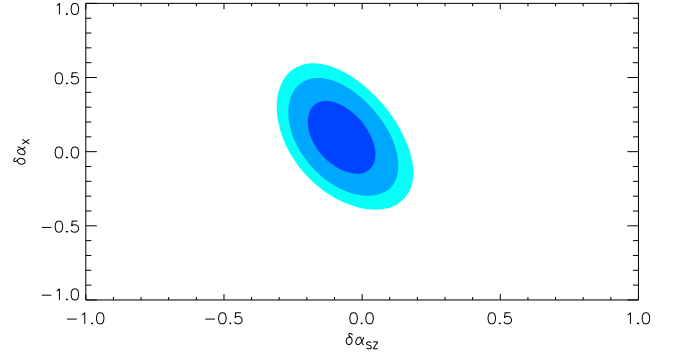}
\caption{Likelihood function as a function of the deviation from scaling law indexes $\delta \alpha_{\rm sz}$ and $\delta \alpha_{\rm x}$ derived from tSZ auto-correlation power spectrum and tSZ-X cross power spectrum.}
\label{likesl}
\end{center}
\end{figure}
In Fig.~\ref{likesl}, we present the constraints from our simulated data. The tSZ-X cross correlation is only sensitive to the sum $\delta \alpha_{\rm x} + \delta \alpha_{\rm sz}$. To distinguish the two indexes, we used the tSZ auto-correlation power spectrum assuming uncertainty levels from \citet{planckSZS}.\\
This measurement is limited by the multipole range accessible for example by {\it Planck} and ROSAT, $\ell \leq 2000$. 
In that case, we obtained $\Delta \left(\delta \alpha_{\rm sz}\right) = 0.10$ and $\Delta \left(\delta \alpha_{\rm sz}\right) = 0.16$ for the scaling law index deviation parameters.
If we add a measurement of tSZ spectrum at $\ell = 3000$, considering tSZ data at higher resolution from SPT-like \citep{shi11} or ACT-like \citep{sie13} experiments, we can reach $\Delta \left(\delta \alpha_{\rm sz}\right) = 0.08$ and $\Delta \left(\delta \alpha_{\rm x}\right) = 0.14$. This constraints on the scaling-law slopes with respect to the cluster mass are competitive with constraints from scaling law dedicated analysis presented in Table.~\ref{tabscal}.\\

\section{Conclusion}

We have presented a complete up-to-date modeling of the tSZ, X-ray and tSZ-X power spectra. We have have carefully studied the sensitivity to cosmological and astrophysical parameters, completing previous analysis on the topic \citep{die03}.
For the variations of  tSZ-X cross correlation, tSZ and X-ray auto-correlation with $H_0$, $\Omega_{\rm m}$ and $\sigma_8$, we have provided accurate analytical fitting formulae in the range $0 < \ell < 10\,000$.\\
Then, we have carefully propagated the uncertainties on the cosmological and scaling law parameters of our modeling to the predicted power spectra, leading to an overall uncertainty on the normalization of the tSZ-X power spectrum by about $47\%$. This result highlights our limited knowledge on this cosmological probes.\\
We note that the contributions to the total uncertainty from clusters scaling relations and cosmological parameters are at same order of magnitude. 
The main contributions to the total uncertainty are produced by the bias $b$, 35\%, and cosmological parameters, 27\%.
This large uncertainties illustrate the importance of an accurate measurement of the tSZ-X cross angular spectrum to set constraints on both cluster properties and cosmological parameters.\\
We stress that most of our modeling parameters act, on the tSZ-X power spectrum, as an overall amplitude factor. This leads to a high degeneracy between those parameters, and thus the tSZ-X cross correlation needs to be used in addition of other probes to break the degeneracies.\\
We note that our prediction cannot be directly be compared to the measurement recently performed by \citet{haj13}, as their measurement concerned the cross-correlation between tSZ maps and X-ray catalogs, and consequently, presents difference dependencies with cosmological parameters.\\

We have predicted the expected signal-to-noise that can be reached using a simulated microwave sky, and ROSAT data for the X-ray sky. We have considered three approaches to extract the tSZ-X cross power spectrum. We demonstrate that in the case of the cross-correlation between a tSZ map and an X-ray map, we reach a signal-to-noise of $31.5$. 
In this case, we can reach a measurement of the tSZ-X cross correlation amplitude at about 3\%, improving the actual constraints (from our knowledge of cosmological parameters and scaling laws) on the predicted spectrum by about a factor of 10.\\

We also study the possibility to constrain the slope of the tSZ and X-ray scaling laws using the shape of the tSZ-X cross spectrum. 
We conclude that extending the tSZ-X measurement up to $\ell=3000$ allows to measure the scaling law indexes at the same level of accuracy than the one presently provided by scaling law dedicated analysis. 
It is worth noting that, with the tSZ-X cross spectrum, we do not have any selection function, we are sensitive to all clusters on the sky weighted by their fluxes.\\

Future experiment at high resolution and high sensitivity for tSZ survey \citep{prism} and X-ray measurements \citep{pre10} will, in the near future, allow to increase the expected constraints. Especially by providing a larger range in multipoles, a higher sensitivity and a larger number of frequency.\\

\label{discon}

\section*{Acknowledgements}
\thanks{We acknowledge the support of the French \emph{Agence Nationale de la Recherche} under grant ANR-11-BD56-015.
We thank E. Pointecouteau for useful discussions related to this work.
This research has made use of the ROSAT all-sky survey data which have been processed at MPE.  
We acknowledge the use of {\tt HEALPix} package \citep{gor05}.
}

\bibliographystyle{aa}
\bibliography{powspec_xsz_th}

\begin{thebibliography}{52}
\expandafter\ifx\csname natexlab\endcsname\relax\def\natexlab#1{#1}\fi

\bibitem[{{Andreon}(2012)}]{and12}
{Andreon}, S. 2012, \aap, 546, A6

\bibitem[{{Arnaud} {et~al.}(2010){Arnaud}, {Pratt}, {Piffaretti},
  {B{\"o}hringer}, {Croston}, \& {Pointecouteau}}]{arn10}
{Arnaud}, M., {Pratt}, G.~W., {Piffaretti}, R., {et~al.} 2010, \aap, 517, A92

\bibitem[{{Benson} {et~al.}(2013){Benson}, {de Haan}, {Dudley}, {Reichardt},
  {Aird}, {Andersson}, {Armstrong}, {Ashby}, {Bautz}, {Bayliss}, {Bazin},
  {Bleem}, {Brodwin}, {Carlstrom}, {Chang}, {Cho}, {Clocchiatti}, {Crawford},
  {Crites}, {Desai}, {Dobbs}, {Foley}, {Forman}, {George}, {Gladders},
  {Gonzalez}, {Halverson}, {Harrington}, {High}, {Holder}, {Holzapfel},
  {Hoover}, {Hrubes}, {Jones}, {Joy}, {Keisler}, {Knox}, {Lee}, {Leitch},
  {Liu}, {Lueker}, {Luong-Van}, {Mantz}, {Marrone}, {McDonald}, {McMahon},
  {Mehl}, {Meyer}, {Mocanu}, {Mohr}, {Montroy}, {Murray}, {Natoli}, {Padin},
  {Plagge}, {Pryke}, {Rest}, {Ruel}, {Ruhl}, {Saliwanchik}, {Saro}, {Sayre},
  {Schaffer}, {Shaw}, {Shirokoff}, {Song}, {Spieler}, {Stalder},
  {Staniszewski}, {Stark}, {Story}, {Stubbs}, {Suhada}, {van Engelen},
  {Vanderlinde}, {Vieira}, {Vikhlinin}, {Williamson}, {Zahn}, \&
  {Zenteno}}]{ben13}
{Benson}, B.~A., {de Haan}, T., {Dudley}, J.~P., {et~al.} 2013, \apj, 763, 147

\bibitem[{{B{\"o}hringer} {et~al.}(2007){B{\"o}hringer}, {Schuecker}, {Pratt},
  {Arnaud}, {Ponman}, {Croston}, {Borgani}, {Bower}, {Briel}, {Collins},
  {Donahue}, {Forman}, {Finoguenov}, {Geller}, {Guzzo}, {Henry}, {Kneissl},
  {Mohr}, {Matsushita}, {Mullis}, {Ohashi}, {Pedersen}, {Pierini}, {Quintana},
  {Raychaudhury}, {Reiprich}, {Romer}, {Rosati}, {Sabirli}, {Temple}, {Viana},
  {Vikhlinin}, {Voit}, \& {Zhang}}]{boh07}
{B{\"o}hringer}, H., {Schuecker}, P., {Pratt}, G.~W., {et~al.} 2007, \aap, 469,
  363

\bibitem[{{Bohringer} {et~al.}(2000){Bohringer}, {Voges}, {Huchra}, {McLean},
  {Giacconi}, {Rosati}, {Burg}, {Mader}, {Schuecker}, {Simic}, {Komossa},
  {Reiprich}, {Retzlaff}, \& {Trumper}}]{boh00}
{Bohringer}, H., {Voges}, W., {Huchra}, J.~P., {et~al.} 2000, VizieR Online
  Data Catalog, 212, 90435

\bibitem[{{Delabrouille} {et~al.}(2013){Delabrouille}, {Betoule}, {Melin},
  {Miville-Desch{\^e}nes}, {Gonzalez-Nuevo}, {Le Jeune}, {Castex}, {de Zotti},
  {Basak}, {Ashdown}, {Aumont}, {Baccigalupi}, {Banday}, {Bernard}, {Bouchet},
  {Clements}, {da Silva}, {Dickinson}, {Dodu}, {Dolag}, {Elsner}, {Fauvet},
  {Fa{\"y}}, {Giardino}, {Leach}, {Lesgourgues}, {Liguori},
  {Mac{\'{\i}}as-P{\'e}rez}, {Massardi}, {Matarrese}, {Mazzotta}, {Montier},
  {Mottet}, {Paladini}, {Partridge}, {Piffaretti}, {Prezeau}, {Prunet},
  {Ricciardi}, {Roman}, {Schaefer}, \& {Toffolatti}}]{psm}
{Delabrouille}, J., {Betoule}, M., {Melin}, J.-B., {et~al.} 2013, \aap, 553,
  A96

\bibitem[{{Diego} {et~al.}(2003){Diego}, {Silk}, \& {Sliwa}}]{die03}
{Diego}, J.~M., {Silk}, J., \& {Sliwa}, W. 2003, \mnras, 346, 940

\bibitem[{{Ebeling} {et~al.}(2000){Ebeling}, {Edge}, {Allen}, {Crawford},
  {Fabian}, \& {Huchra}}]{ebe00}
{Ebeling}, H., {Edge}, A.~C., {Allen}, S.~W., {et~al.} 2000, VizieR Online Data
  Catalog, 731, 80333

\bibitem[{{Ebeling} {et~al.}(2001){Ebeling}, {Edge}, \& {Henry}}]{ebe01}
{Ebeling}, H., {Edge}, A.~C., \& {Henry}, J.~P. 2001, \apj, 553, 668

\bibitem[{{Evrard} {et~al.}(2002){Evrard}, {MacFarland}, {Couchman}, {Colberg},
  {Yoshida}, {White}, {Jenkins}, {Frenk}, {Pearce}, {Peacock}, \&
  {Thomas}}]{evr02}
{Evrard}, A.~E., {MacFarland}, T.~J., {Couchman}, H.~M.~P., {et~al.} 2002,
  \apj, 573, 7

\bibitem[{{Fixsen} {et~al.}(1998){Fixsen}, {Dwek}, {Mather}, {Bennett}, \&
  {Shafer}}]{fix98}
{Fixsen}, D.~J., {Dwek}, E., {Mather}, J.~C., {Bennett}, C.~L., \& {Shafer},
  R.~A. 1998, \apj, 508, 123

\bibitem[{{Freyberg} {et~al.}(1992){Freyberg}, {Snowden}, {Plucinsky}, \&
  {Schmitt}}]{fre92}
{Freyberg}, M.~J., {Snowden}, S.~L., {Plucinsky}, P.~P., \& {Schmitt},
  J.~H.~M.~M. 1992, in Astronomische Gesellschaft Abstract Series, Vol.~7,
  Astronomische Gesellschaft Abstract Series, ed. G.~{Klare}, 26

\bibitem[{{Gispert} {et~al.}(2000){Gispert}, {Lagache}, \& {Puget}}]{gis00}
{Gispert}, R., {Lagache}, G., \& {Puget}, J.~L. 2000, \aap, 360, 1

\bibitem[{{G{\'o}rski} {et~al.}(2005){G{\'o}rski}, {Hivon}, {Banday},
  {Wandelt}, {Hansen}, {Reinecke}, \& {Bartelmann}}]{gor05}
{G{\'o}rski}, K.~M., {Hivon}, E., {Banday}, A.~J., {et~al.} 2005, \apj, 622,
  759

\bibitem[{{Hajian} {et~al.}(2013){Hajian}, {Battaglia}, {Spergel}, {Bond},
  {Pfrommer}, \& {Sievers}}]{haj13}
{Hajian}, A., {Battaglia}, N., {Spergel}, D.~N., {et~al.} 2013, ArXiv e-prints

\bibitem[{{Hurier} {et~al.}(2013{\natexlab{a}}){Hurier}, {Aghanim}, {Douspis},
  \& {Pointecouteau}}]{hur13b}
{Hurier}, G., {Aghanim}, N., {Douspis}, M., \& {Pointecouteau}, E.
  2013{\natexlab{a}}, ArXiv e-prints

\bibitem[{{Hurier} {et~al.}(2013{\natexlab{b}}){Hurier},
  {Mac{\'{\i}}as-P{\'e}rez}, \& {Hildebrandt}}]{hur13a}
{Hurier}, G., {Mac{\'{\i}}as-P{\'e}rez}, J.~F., \& {Hildebrandt}, S.
  2013{\natexlab{b}}, \aap, 558, A118

\bibitem[{{Komatsu} \& {Kitayama}(1999)}]{kom99}
{Komatsu}, E. \& {Kitayama}, T. 1999, \apjl, 526, L1

\bibitem[{{Komatsu} \& {Seljak}(2001)}]{kom01}
{Komatsu}, E. \& {Seljak}, U. 2001, \mnras, 327, 1353

\bibitem[{{Komatsu} \& {Seljak}(2002)}]{kom02}
{Komatsu}, E. \& {Seljak}, U. 2002, \mnras, 336, 1256

\bibitem[{{Krumpe} {et~al.}(2010){Krumpe}, {Miyaji}, \& {Coil}}]{kru10}
{Krumpe}, M., {Miyaji}, T., \& {Coil}, A.~L. 2010, \apj, 713, 558

\bibitem[{{Krumpe} {et~al.}(2012){Krumpe}, {Miyaji}, {Coil}, \&
  {Aceves}}]{kru12}
{Krumpe}, M., {Miyaji}, T., {Coil}, A.~L., \& {Aceves}, H. 2012, \apj, 746, 1

\bibitem[{{Lesgourgues}(2011)}]{les11}
{Lesgourgues}, J. 2011, ArXiv e-prints

\bibitem[{{Marriage} {et~al.}(2011){Marriage}, {Acquaviva}, {Ade}, {Aguirre},
  {Amiri}, {Appel}, {Barrientos}, {Battistelli}, {Bond}, {Brown}, {Burger},
  {Chervenak}, {Das}, {Devlin}, {Dicker}, {Bertrand Doriese}, {Dunkley},
  {D{\"u}nner}, {Essinger-Hileman}, {Fisher}, {Fowler}, {Hajian}, {Halpern},
  {Hasselfield}, {Hern{\'a}ndez-Monteagudo}, {Hilton}, {Hilton}, {Hincks},
  {Hlozek}, {Huffenberger}, {Handel Hughes}, {Hughes}, {Infante}, {Irwin},
  {Baptiste Juin}, {Kaul}, {Klein}, {Kosowsky}, {Lau}, {Limon}, {Lin},
  {Lupton}, {Marsden}, {Martocci}, {Mauskopf}, {Menanteau}, {Moodley},
  {Moseley}, {Netterfield}, {Niemack}, {Nolta}, {Page}, {Parker}, {Partridge},
  {Quintana}, {Reese}, {Reid}, {Sehgal}, {Sherwin}, {Sievers}, {Spergel},
  {Staggs}, {Swetz}, {Switzer}, {Thornton}, {Trac}, {Tucker}, {Warne},
  {Wilson}, {Wollack}, \& {Zhao}}]{mar11}
{Marriage}, T.~A., {Acquaviva}, V., {Ade}, P.~A.~R., {et~al.} 2011, \apj, 737,
  61

\bibitem[{{Mewe} {et~al.}(1985){Mewe}, {Gronenschild}, \& {van den
  Oord}}]{mew85}
{Mewe}, R., {Gronenschild}, E.~H.~B.~M., \& {van den Oord}, G.~H.~J. 1985,
  \aaps, 62, 197

\bibitem[{{Miyaji} {et~al.}(2011){Miyaji}, {Krumpe}, {Coil}, \&
  {Aceves}}]{miy11}
{Miyaji}, T., {Krumpe}, M., {Coil}, A.~L., \& {Aceves}, H. 2011, \apj, 726, 83

\bibitem[{{Mo} \& {White}(1996)}]{mo96}
{Mo}, H.~J. \& {White}, S.~D.~M. 1996, \mnras, 282, 347

\bibitem[{{Nagai} {et~al.}(2007){Nagai}, {Kravtsov}, \& {Vikhlinin}}]{nag07}
{Nagai}, D., {Kravtsov}, A.~V., \& {Vikhlinin}, A. 2007, \apj, 668, 1

\bibitem[{{Navarro} {et~al.}(1997){Navarro}, {Frenk}, \& {White}}]{nav97}
{Navarro}, J.~F., {Frenk}, C.~S., \& {White}, S.~D.~M. 1997, \apj, 490, 493

\bibitem[{{Piffaretti} \& {Valdarnini}(2008)}]{pif08}
{Piffaretti}, R. \& {Valdarnini}, R. 2008, \aap, 491, 71

\bibitem[{{Planck Collaboration} {et~al.}(2013{\natexlab{a}}){Planck
  Collaboration}, {Ade}, {Aghanim}, {Armitage-Caplan}, {Arnaud}, {Ashdown},
  {Atrio-Barandela}, {Aumont}, {Baccigalupi}, {Banday}, \& et~al.}]{PlanckOVER}
{Planck Collaboration}, {Ade}, P.~A.~R., {Aghanim}, N., {et~al.}
  2013{\natexlab{a}}, ArXiv e-prints

\bibitem[{{Planck Collaboration} {et~al.}(2013{\natexlab{b}}){Planck
  Collaboration}, {Ade}, {Aghanim}, {Armitage-Caplan}, {Arnaud}, {Ashdown},
  {Atrio-Barandela}, {Aumont}, {Baccigalupi}, {Banday}, \& et~al.}]{PlanckSZC}
{Planck Collaboration}, {Ade}, P.~A.~R., {Aghanim}, N., {et~al.}
  2013{\natexlab{b}}, ArXiv e-prints

\bibitem[{{Planck Collaboration} {et~al.}(2013{\natexlab{c}}){Planck
  Collaboration}, {Ade}, {Aghanim}, {Armitage-Caplan}, {Arnaud}, {Ashdown},
  {Atrio-Barandela}, {Aumont}, {Baccigalupi}, {Banday}, \& et~al.}]{planckPAR}
{Planck Collaboration}, {Ade}, P.~A.~R., {Aghanim}, N., {et~al.}
  2013{\natexlab{c}}, ArXiv e-prints

\bibitem[{{Planck Collaboration} {et~al.}(2011{\natexlab{a}}){Planck
  Collaboration}, {Ade}, {Aghanim}, {Arnaud}, {Ashdown}, {Aumont},
  {Baccigalupi}, {Balbi}, {Banday}, {Barreiro}, \& et~al.}]{planckSZS}
{Planck Collaboration}, {Ade}, P.~A.~R., {Aghanim}, N., {et~al.}
  2011{\natexlab{a}}, \aap, 536, A8

\bibitem[{{Planck Collaboration} {et~al.}(2011{\natexlab{b}}){Planck
  Collaboration}, {Ade}, {Aghanim}, {Arnaud}, {Ashdown}, {Aumont},
  {Baccigalupi}, {Balbi}, {Banday}, {Barreiro}, \& et~al.}]{planckSL}
{Planck Collaboration}, {Ade}, P.~A.~R., {Aghanim}, N., {et~al.}
  2011{\natexlab{b}}, \aap, 536, A11

\bibitem[{{Planck Collaboration early VIII}(2011)}]{planckESZ}
{Planck Collaboration early VIII}. 2011, \aap, 536, A8

\bibitem[{{Planck Collaboration results XXIX}(2013)}]{PlanckPSZ}
{Planck Collaboration results XXIX}. 2013, e-prints ArXiv: 1303.5089

\bibitem[{{Predehl} {et~al.}(2010){Predehl}, {Andritschke}, {B{\"o}hringer},
  {Bornemann}, {Br{\"a}uninger}, {Brunner}, {Brusa}, {Burkert}, {Burwitz},
  {Cappelluti}, {Churazov}, {Dennerl}, {Eder}, {Elbs}, {Freyberg}, {Friedrich},
  {F{\"u}rmetz}, {Gaida}, {H{\"a}lker}, {Hartner}, {Hasinger}, {Hermann},
  {Huber}, {Kendziorra}, {von Kienlin}, {Kink}, {Kreykenbohm}, {Lamer},
  {Lapchov}, {Lehmann}, {Meidinger}, {Mican}, {Mohr}, {M{\"u}hlegger},
  {M{\"u}ller}, {Nandra}, {Pavlinsky}, {Pfeffermann}, {Reiprich}, {Robrade},
  {Roh{\'e}}, {Santangelo}, {Sch{\"a}chner}, {Schanz}, {Schmid}, {Schmitt},
  {Schreib}, {Schrey}, {Schwope}, {Steinmetz}, {Str{\"u}der}, {Sunyaev},
  {Tenzer}, {Tiedemann}, {Vongehr}, \& {Wilms}}]{pre10}
{Predehl}, P., {Andritschke}, R., {B{\"o}hringer}, H., {et~al.} 2010, in
  Society of Photo-Optical Instrumentation Engineers (SPIE) Conference Series,
  Vol. 7732, Society of Photo-Optical Instrumentation Engineers (SPIE)
  Conference Series

\bibitem[{{PRISM Collaboration} {et~al.}(2013){PRISM Collaboration}, {Andre},
  {Baccigalupi}, {Barbosa}, {Bartlett}, {Bartolo}, {Battistelli}, {Battye},
  {Bendo}, {Bernard}, {Bersanelli}, {Bethermin}, {Bielewicz}, {Bonaldi},
  {Bouchet}, {Boulanger}, {Brand}, {Bucher}, {Burigana}, {Cai}, {Casasola},
  {Castex}, {Challinor}, {Chluba}, {Colafrancesco}, {Cuttaia}, {D'Alessandro},
  {Davis}, {de Avillez}, {de Bernardis}, {de Petris}, {de Rosa}, {de Zotti},
  {Delabrouille}, {Dickinson}, {Diego}, {Falgarone}, {Ferreira}, {Ferriere},
  {Finelli}, {Fletcher}, {Fuller}, {Galli}, {Ganga}, {Garcia-Bellido},
  {Ghribi}, {Gonzalez-Nuevo}, {Grainge}, {Gruppuso}, {Hall},
  {Hernandez-Monteagudo}, {Jackson}, {Jaffe}, {Khatri}, {Lamagna}, {Lattanzi},
  {Leahy}, {Liguori}, {Liuzzo}, {Lopez-Caniego}, {Macias-Perez}, {Maffei},
  {Maino}, {Masi}, {Mangilli}, {Massardi}, {Matarrese}, {Melchiorri}, {Melin},
  {Mennella}, {Mignano}, {Miville-Deschenes}, {Nati}, {Natoli}, {Negrello},
  {Noviello}, {Paci}, {Paladino}, {Paoletti}, {Perrotta}, {Piacentini}, {Piat},
  {Piccirillo}, {Pisano}, {Polenta}, {Ricciardi}, {Roman}, {Rubino-Martin},
  {Salatino}, {Schillaci}, {Shellard}, {Silk}, {Stompor}, {Sunyaev}, {Tartari},
  {Terenzi}, {Toffolatti}, {Tomasi}, {Trombetti}, {Tucci}, {Van Tent}, {Verde},
  {Wandelt}, \& {Withington}}]{prism}
{PRISM Collaboration}, {Andre}, P., {Baccigalupi}, C., {et~al.} 2013, ArXiv
  e-prints

\bibitem[{{Puget} {et~al.}(1996){Puget}, {Abergel}, {Bernard}, {Boulanger},
  {Burton}, {Desert}, \& {Hartmann}}]{pug96}
{Puget}, J.-L., {Abergel}, A., {Bernard}, J.-P., {et~al.} 1996, \aap, 308, L5

\bibitem[{{Reichardt} {et~al.}(2013){Reichardt}, {Stalder}, {Bleem}, {Montroy},
  {Aird}, {Andersson}, {Armstrong}, {Ashby}, {Bautz}, {Bayliss}, {Bazin},
  {Benson}, {Brodwin}, {Carlstrom}, {Chang}, {Cho}, {Clocchiatti}, {Crawford},
  {Crites}, {de Haan}, {Desai}, {Dobbs}, {Dudley}, {Foley}, {Forman}, {George},
  {Gladders}, {Gonzalez}, {Halverson}, {Harrington}, {High}, {Holder},
  {Holzapfel}, {Hoover}, {Hrubes}, {Jones}, {Joy}, {Keisler}, {Knox}, {Lee},
  {Leitch}, {Liu}, {Lueker}, {Luong-Van}, {Mantz}, {Marrone}, {McDonald},
  {McMahon}, {Mehl}, {Meyer}, {Mocanu}, {Mohr}, {Murray}, {Natoli}, {Padin},
  {Plagge}, {Pryke}, {Rest}, {Ruel}, {Ruhl}, {Saliwanchik}, {Saro}, {Sayre},
  {Schaffer}, {Shaw}, {Shirokoff}, {Song}, {Spieler}, {Staniszewski}, {Stark},
  {Story}, {Stubbs}, {{\v S}uhada}, {van Engelen}, {Vanderlinde}, {Vieira},
  {Vikhlinin}, {Williamson}, {Zahn}, \& {Zenteno}}]{rei13}
{Reichardt}, C.~L., {Stalder}, B., {Bleem}, L.~E., {et~al.} 2013, \apj, 763,
  127

\bibitem[{{Remazeilles} {et~al.}(2011){Remazeilles}, {Delabrouille}, \&
  {Cardoso}}]{rem11}
{Remazeilles}, M., {Delabrouille}, J., \& {Cardoso}, J.-F. 2011, \mnras, 410,
  2481

\bibitem[{{Sehgal} {et~al.}(2011){Sehgal}, {Trac}, {Acquaviva}, {Ade},
  {Aguirre}, {Amiri}, {Appel}, {Barrientos}, {Battistelli}, {Bond}, {Brown},
  {Burger}, {Chervenak}, {Das}, {Devlin}, {Dicker}, {Bertrand Doriese},
  {Dunkley}, {D{\"u}nner}, {Essinger-Hileman}, {Fisher}, {Fowler}, {Hajian},
  {Halpern}, {Hasselfield}, {Hern{\'a}ndez-Monteagudo}, {Hilton}, {Hilton},
  {Hincks}, {Hlozek}, {Holtz}, {Huffenberger}, {Hughes}, {Hughes}, {Infante},
  {Irwin}, {Jones}, {Baptiste Juin}, {Klein}, {Kosowsky}, {Lau}, {Limon},
  {Lin}, {Lupton}, {Marriage}, {Marsden}, {Martocci}, {Mauskopf}, {Menanteau},
  {Moodley}, {Moseley}, {Netterfield}, {Niemack}, {Nolta}, {Page}, {Parker},
  {Partridge}, {Reid}, {Sherwin}, {Sievers}, {Spergel}, {Staggs}, {Swetz},
  {Switzer}, {Thornton}, {Tucker}, {Warne}, {Wollack}, \& {Zhao}}]{seh11}
{Sehgal}, N., {Trac}, H., {Acquaviva}, V., {et~al.} 2011, \apj, 732, 44

\bibitem[{{Shirokoff} {et~al.}(2011){Shirokoff}, {Reichardt}, {Shaw}, {Millea},
  {Ade}, {Aird}, {Benson}, {Bleem}, {Carlstrom}, {Chang}, {Cho}, {Crawford},
  {Crites}, {de Haan}, {Dobbs}, {Dudley}, {George}, {Halverson}, {Holder},
  {Holzapfel}, {Hrubes}, {Joy}, {Keisler}, {Knox}, {Lee}, {Leitch}, {Lueker},
  {Luong-Van}, {McMahon}, {Mehl}, {Meyer}, {Mohr}, {Montroy}, {Padin},
  {Plagge}, {Pryke}, {Ruhl}, {Schaffer}, {Spieler}, {Staniszewski}, {Stark},
  {Story}, {Vanderlinde}, {Vieira}, {Williamson}, \& {Zahn}}]{shi11}
{Shirokoff}, E., {Reichardt}, C.~L., {Shaw}, L., {et~al.} 2011, \apj, 736, 61

\bibitem[{{Sievers} {et~al.}(2013){Sievers}, {Hlozek}, {Nolta}, {Acquaviva},
  {Addison}, {Ade}, {Aguirre}, {Amiri}, {Appel}, {Barrientos}, {Battistelli},
  {Battaglia}, {Bond}, {Brown}, {Burger}, {Calabrese}, {Chervenak}, {Crichton},
  {Das}, {Devlin}, {Dicker}, {Bertrand Doriese}, {Dunkley}, {D{\"u}nner},
  {Essinger-Hileman}, {Faber}, {Fisher}, {Fowler}, {Gallardo}, {Gordon},
  {Gralla}, {Hajian}, {Halpern}, {Hasselfield}, {Hern{\'a}ndez-Monteagudo},
  {Hill}, {Hilton}, {Hilton}, {Hincks}, {Holtz}, {Huffenberger}, {Hughes},
  {Hughes}, {Infante}, {Irwin}, {Jacobson}, {Johnstone}, {Baptiste Juin},
  {Kaul}, {Klein}, {Kosowsky}, {Lau}, {Limon}, {Lin}, {Louis}, {Lupton},
  {Marriage}, {Marsden}, {Martocci}, {Mauskopf}, {McLaren}, {Menanteau},
  {Moodley}, {Moseley}, {Netterfield}, {Niemack}, {Page}, {Page}, {Parker},
  {Partridge}, {Plimpton}, {Quintana}, {Reese}, {Reid}, {Rojas}, {Sehgal},
  {Sherwin}, {Schmitt}, {Spergel}, {Staggs}, {Stryzak}, {Swetz}, {Switzer},
  {Thornton}, {Trac}, {Tucker}, {Uehara}, {Visnjic}, {Warne}, {Wilson},
  {Wollack}, {Zhao}, \& {Zunckel}}]{sie13}
{Sievers}, J.~L., {Hlozek}, R.~A., {Nolta}, M.~R., {et~al.} 2013, \jcap, 10, 60

\bibitem[{{Sunyaev} \& {Zeldovich}(1969)}]{sun69}
{Sunyaev}, R.~A. \& {Zeldovich}, Y.~B. 1969, \nat, 223, 721

\bibitem[{{Sunyaev} \& {Zeldovich}(1972)}]{sun72}
{Sunyaev}, R.~A. \& {Zeldovich}, Y.~B. 1972, Comments on Astrophysics and Space
  Physics, 4, 173

\bibitem[{{Taburet} {et~al.}(2011){Taburet}, {Hern{\'a}ndez-Monteagudo},
  {Aghanim}, {Douspis}, \& {Sunyaev}}]{tab11}
{Taburet}, N., {Hern{\'a}ndez-Monteagudo}, C., {Aghanim}, N., {Douspis}, M., \&
  {Sunyaev}, R.~A. 2011, \mnras, 418, 2207

\bibitem[{{Tinker} {et~al.}(2008){Tinker}, {Kravtsov}, {Klypin}, {Abazajian},
  {Warren}, {Yepes}, {Gottl{\"o}ber}, \& {Holz}}]{tin08}
{Tinker}, J., {Kravtsov}, A.~V., {Klypin}, A., {et~al.} 2008, \apj, 688, 709

\bibitem[{{Tristram} {et~al.}(2005){Tristram}, {Mac{\'{\i}}as-P{\'e}rez},
  {Renault}, \& {Santos}}]{tri05}
{Tristram}, M., {Mac{\'{\i}}as-P{\'e}rez}, J.~F., {Renault}, C., \& {Santos},
  D. 2005, \mnras, 358, 833

\bibitem[{{Vanderlinde} {et~al.}(2010){Vanderlinde}, {Crawford}, {de Haan},
  {Dudley}, {Shaw}, {Ade}, {Aird}, {Benson}, {Bleem}, {Brodwin}, {Carlstrom},
  {Chang}, {Crites}, {Desai}, {Dobbs}, {Foley}, {George}, {Gladders}, {Hall},
  {Halverson}, {High}, {Holder}, {Holzapfel}, {Hrubes}, {Joy}, {Keisler},
  {Knox}, {Lee}, {Leitch}, {Loehr}, {Lueker}, {Marrone}, {McMahon}, {Mehl},
  {Meyer}, {Mohr}, {Montroy}, {Ngeow}, {Padin}, {Plagge}, {Pryke}, {Reichardt},
  {Rest}, {Ruel}, {Ruhl}, {Schaffer}, {Shirokoff}, {Song}, {Spieler},
  {Stalder}, {Staniszewski}, {Stark}, {Stubbs}, {van Engelen}, {Vieira},
  {Williamson}, {Yang}, {Zahn}, \& {Zenteno}}]{van10}
{Vanderlinde}, K., {Crawford}, T.~M., {de Haan}, T., {et~al.} 2010, \apj, 722,
  1180

\bibitem[{{Voges} {et~al.}(1999){Voges}, {Aschenbach}, {Boller},
  {Br{\"a}uninger}, {Briel}, {Burkert}, {Dennerl}, {Englhauser}, {Gruber},
  {Haberl}, {Hartner}, {Hasinger}, {K{\"u}rster}, {Pfeffermann}, {Pietsch},
  {Predehl}, {Rosso}, {Schmitt}, {Tr{\"u}mper}, \& {Zimmermann}}]{RASS}
{Voges}, W., {Aschenbach}, B., {Boller}, T., {et~al.} 1999, \aap, 349, 389

\end{thebibliography}

\end{document}